\documentclass[review]{elsarticle}

\usepackage{lineno,hyperref}
\modulolinenumbers[5]
\usepackage{epsfig}
\usepackage{amsmath}
\usepackage{amssymb}
\usepackage{mathptmx}
\usepackage{graphicx}
\usepackage{xfrac}
\usepackage{comment}
\usepackage{citesort}
\usepackage{epstopdf}
\usepackage{color}
\usepackage{subfig}

\newtheorem{definition}{Definition}
\newtheorem{thm}{Theorem}

\newdefinition{rmk}{Remark}
\newproof{pf}{Proof}

\journal{Robotics and Autonomous Systems}

\title{Collaborative Visual Area Coverage\tnoteref{mytitlenote}}

\tnotetext[mytitlenote]{This work has received funding from the European Union's Horizon 2020 Research and Innovation Programme under the Grant Agreement No.644128, AEROWORKS. A shorter version, without the stability--notion and subsequent proof has been submitted for possible inclusion at the ICRA 2017 proceedings.}

\author{Sotiris Papatheodorou}

\author{Anthony Tzes}

\author{Yiannis Stergiopoulos\fnref{myfootnote}}

\fntext[myfootnote]{The authors are with the Electrical \& Computer Engineering Department, University of Patras, Rio, Achaia 26500, Greece. Corresponding author's email: \small\tt{tzes@ece.upatras.gr}}

\begin{document}
	\begin{frontmatter}

	\begin{abstract}
	
	This article examines the problem of visual area coverage by a network of Mobile Aerial Agents (MAAs). Each MAA is assumed to be equipped with a downwards facing camera with a conical field of view which covers all points within a circle on the ground. The diameter of that circle is proportional to the altitude of the MAA, whereas the quality of the covered area decreases with the altitude. A distributed control law that maximizes a joint coverage-quality criterion by adjusting the MAAs' spatial coordinates is developed. The effectiveness of the proposed control scheme is evaluated through simulation studies.
	\end{abstract}
	
	
	\end{frontmatter}
	
	\linenumbers

%%%%%%%%%%%%%%%%%%%%%%%%%%%%%%%%%%%%%%%%%%%%%%%%%%%%%%%%%%%%%%%%%%%%%%%%%%%%%%%%
\section{Introduction}
	Area coverage over a planar region by ground agents has been studied extensively when the sensing patterns of the agents are circular \cite{Cortes_ESAIMCOCV05,Pimenta_CDC08}. Most of these techniques are based on a Voronoi or similar partitioning~\cite{Stergiopoulos_IETCTA10,Arslan_ICRA2016,Nguyen_ISIC2016} of the region of interest and use distributed optimization, model predictive control~\cite{Nguyen_MED2016,Mohseni_IEEESJ2016} or game theory~\cite{Ramaswamy_ACC2016} among other techniques. There is also significant work concerning arbitrary sensing patterns~\cite{Stergiopoulos_Automatica13,Kantaros_Automatica15,Panagou_IEEETACCNS2016} avoiding the usage of Voronoi partitioning~\cite{Stergiopoulos_ICRA14,Bakolas_SCL2016}. Both convex and non-convex domains have been examined~\cite{Stergiopoulos_IEEETAC15,Alitappeh_SC2016}.
	
	Many algorithms have been developed for mapping by  MAAs~\cite{Renzaglia_IJRR12,Breitenmoser_IROS10,Thanou_ISCCSP14,Torres_ESA2016} relying mostly in Voronoi-based tessellations or path--planning. Extensive work has also been done in area monitoring by MAAs equipped with cameras~\cite{Schwager_ICRA2009,Schwager_IEEE2011}. In these pioneering research efforts, there is no maximum allowable height that can be reached by the  MAAs and the case where there is overlapping of their covered areas is considered an advantage as opposed to the same area viewed by a single camera. There are also studies on the connectivity and energy consumption of MAA networks~\cite{Yanmaz_ICC2012,Messous_WCNC2016}.
	
	In this paper the persistent coverage problem of a convex planar region by a network of MAAs is considered. The MAAs are assumed to have downwards facing visual sensors with a conical field of view, thus creating a circular sensing footprint. The covered area as well as the coverage quality of that area are dependent on the altitude of each MAA. MAAs at higher altitudes cover more area but the coverage quality is smaller compared to MAAs at lower altitudes. A partitioning scheme of the sensed region, similar to~\cite{Stergiopoulos_ICRA14}, is employed and a gradient based control law is developed. This control law leads the network to a locally optimal configuration with respect to a combined coverage-quality criterion, while also guaranteeing that the MAAs remain within a desired range of altitudes. The main contribution of this work is the guarantee it offers that all MAAs will remain within a predefined altitude range. In addition to that, overlapping between the sensed regions of different  MAAs is avoided if possible, in contrast to previous works which consider it an advantage.
	
	The problem statement and the joint coverage--quality criterion are presented in Section \ref{section:problem_statement}. The chosen quality function is defined in Section \ref{section:coverage_quality} and the resulting sensed space partitioning scheme in Section \ref{section:partitioning}. The distributed control law is derived and its most notable properties explained in Section \ref{section:distributed_control_law}. The stability of the altitude control law and its property to restrict the nodes' altitude is examined in Section \ref{section:stability}. Simulation studies highlighting the efficiency of the proposed control law are provided in Section \ref{section:simulation_studies} followed by concluding remarks.

%%%%%%%%%%%%%%%%%%%%%%%%%%%%%%%%%%%%%%%%%%%%%%%%%%%%%%%%%%%%%%%%%%%%%%%%%%%%%%%%
\section{Problem Statement}
\label{section:problem_statement}

	Let $\Omega \subset \mathbb{R}^2$ be a compact convex region under surveillance. We assume a swarm of $n$  MAAs, each positioned at the spatial coordinates $X_i = \left[ x_i~y_i~z_i \right]^T, ~i \in I_n$, where $I_n = \left\{ 1, \dots ,n \right\}$. We also define the vector $q_i = [ x_i ~ y_i ]^T, ~q_i \in \Omega$ to note the projection of the center of each MAA on the ground. The minimum and maximum altitudes each MAA can fly to are $z_i^{\min}$ and $z_i^{\max}$ respectively, thus $z_i \in [z_i^{\min}, ~z_i^{\max}], ~i \in I_n$. It is also assumed that $z_i^{\min} > 0, ~\forall i \in I_n$, since setting the minimum altitude to zero could potentially cause some MAAs to crash.
	
	The simplified MAA's kinodynamic model is 
	\begin{eqnarray}
		\nonumber
		\dot{q}_i &=& u_{i,q},~~q_i \in \Omega, ~u_{i,q} \in \mathbb{R}^2,\\
		\dot{z}_i &=& u_{i,z},~~z_i \in [z_i^{\min} ~,~ z_i^{\max}], ~u_{i,z}\in \mathbb{R}.
		\label{kinematics}
	\end{eqnarray}
	where $\left[u_{i,q}, u_{i,z}\right]$ is the corresponding `thrust' control input for each MAA~(node). The minimum altitude $z_i^{\min}$ is used to ensure the MAAs will fly above ground obstacles, whereas the maximum altitude $z_i^{\max}$ guarantees that they will not fly out of range of their base station. In the sequel, all MAAs are assumed to have common minimum $z^{\min}$ and maximum $z^{\max}$ altitudes.

\begin{figure}[htb]
		\centering
		\includegraphics[width=0.5\textwidth]{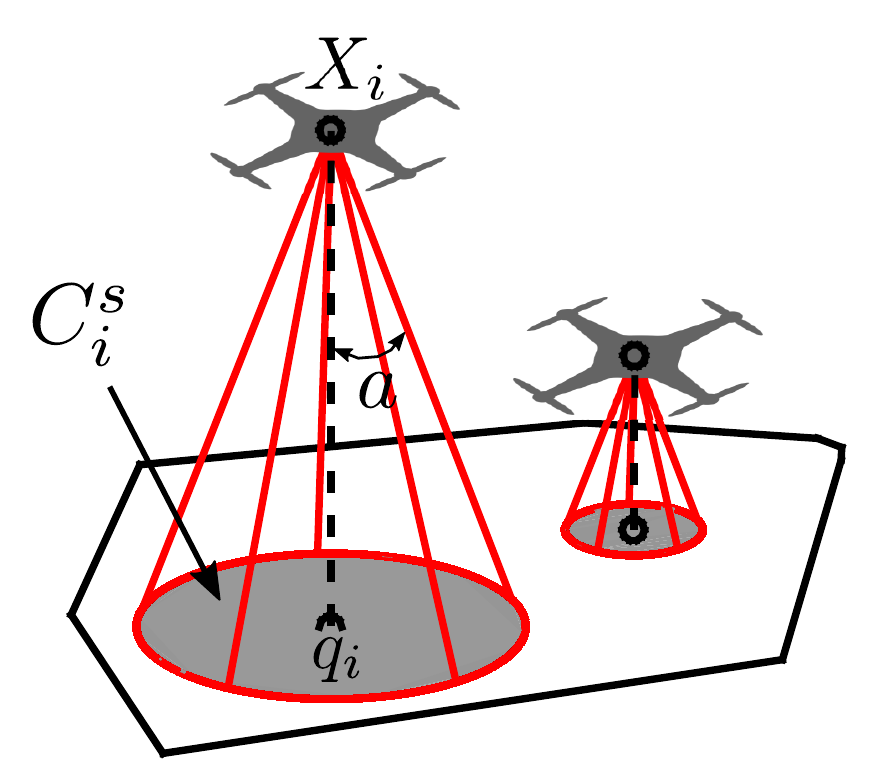}
		\caption{MAA-visual area coverage concept}
		\label{fig:problem_statement}
	\end{figure}

	As far as the sensing performance of the MAAs ~(nodes) is concerned, all members are assumed to be equipped with identical downwards pointing sensors with conic sensing patterns. Thus the region of $\Omega$ sensed by each node is a disk defined as
	\begin{equation}
		C_{i}^{s}(X_i,a) = \left\{ q \in \Omega: \parallel q-q_i\parallel \leq z_i~ \tan a \right\},~i=1,\ldots,n,
		\label{sensing}
	\end{equation}
	where $a$ is half the angle of the sensing cone. As shown in Figure \ref{fig:problem_statement}, the higher the altitude of an MAA, the larger the area of $\Omega$ surveyed by its sensor. 
	
	The coverage quality of each node is a function $f(z_i)\colon [z^{\min}, ~z^{\max}] \rightarrow [0, 1]$ which is dependent on the node's altitude constraints $z^{\min}$ and $z^{\max}$. The coverage quality of node $i$ is assumed to be uniform throughout its sensed region $C_{i}^{s}$. The higher the value of $f(z_i)$, the better the coverage quality. It is assumed that as the altitude of a node increases, the visual quality of its sensed area decreases. The exact definition and properties of $f(z_i)$ are presented in Section \ref{section:coverage_quality}. 

	For each point $q \in \Omega$, an importance weight is assigned via the space density function $\phi \colon \Omega \rightarrow \mathbb{R}^+$, encapsulating any a priori information regarding the region of interest. Thus the coverage-quality objective is
	\begin{equation}
	\mathcal{H} \stackrel{\triangle}{=} \int_{\Omega} \max_{i \in I_n} f(z_i) ~\phi(q) ~dq.
	\end{equation}
	In the sequel, we assume $\phi(q) = 1, ~\forall q \in \Omega$ but the expressions can be easily altered to take into account any a priori weight function.

%%%%%%%%%%%%%%%%%%%%%%%%%%%%%%%%%%%%%%%%%
\section{Coverage quality function}
\label{section:coverage_quality}
%%%%%%%%%%%%%%%%%%%%%%%%%%%%%%%%%%%%%%%%%
	A uniform coverage quality throughout the sensed region $C_i^s$ can be used to model downward facing cameras~\cite{DiFranco_JIRS2016,Avellar_S2015} that provide uniform quality in the whole image. The uniform coverage quality function $f(z_i)\colon [z^{\min}, ~z^{\max}] \rightarrow [0, 1]$ was chosen to be
	\begin{equation*}
	f(z_i) = \left \{
	\begin{aligned}
		&~\frac{\left( \left( z_i - z^{\min} \right)^2 - \left( z^{\max} - z^{\min} \right)^2 \right)^2}{\left( z^{\max} - z^{\min} \right)^4}, & ~q \in C_i^s\\
		&~0, & ~q \notin C_i^s
	\end{aligned}
	\right.\\
	\end{equation*}

\begin{figure}[htb]
		\centering
		\includegraphics[width=0.49\textwidth]{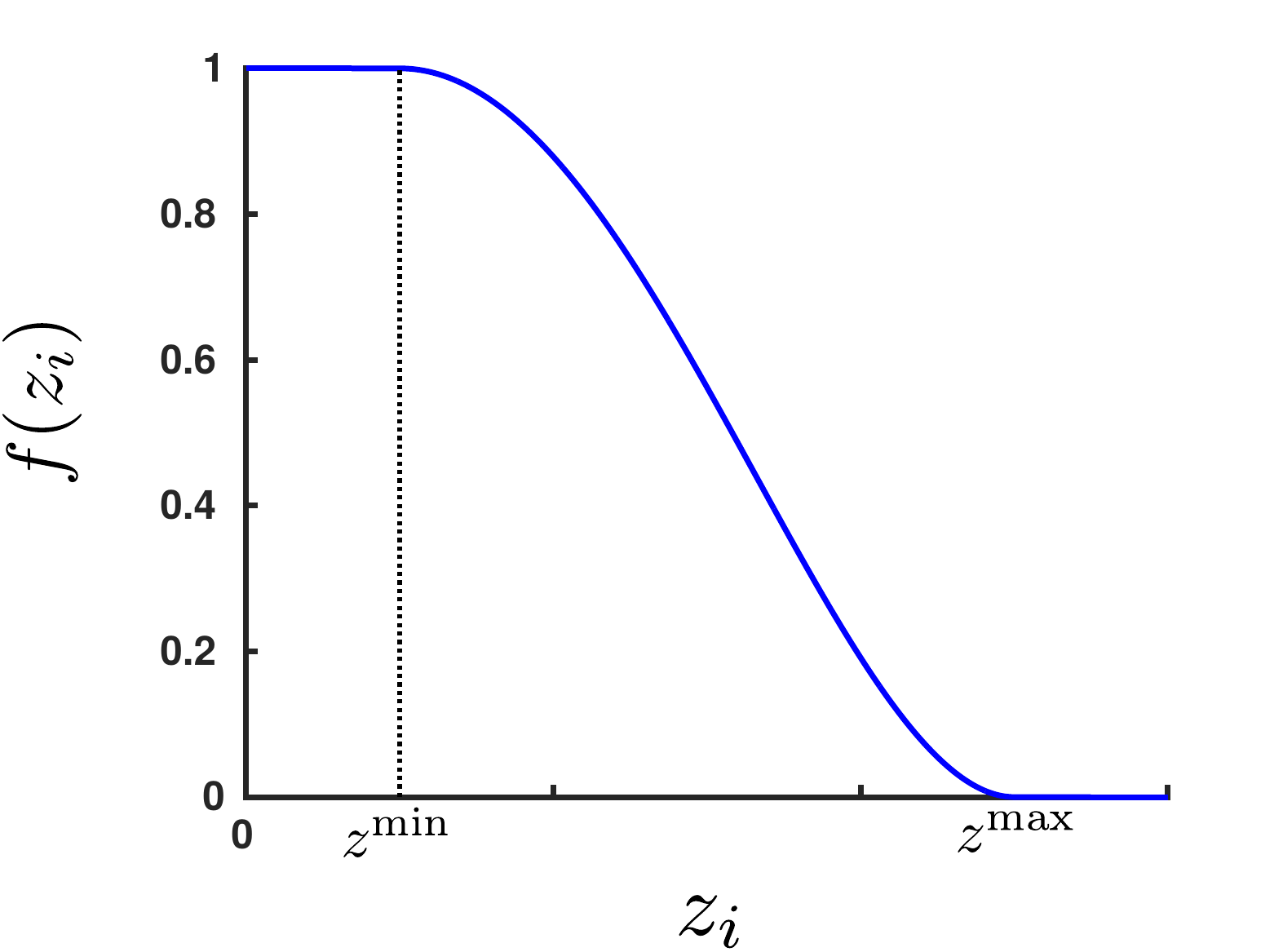}
		\includegraphics[width=0.49\textwidth]{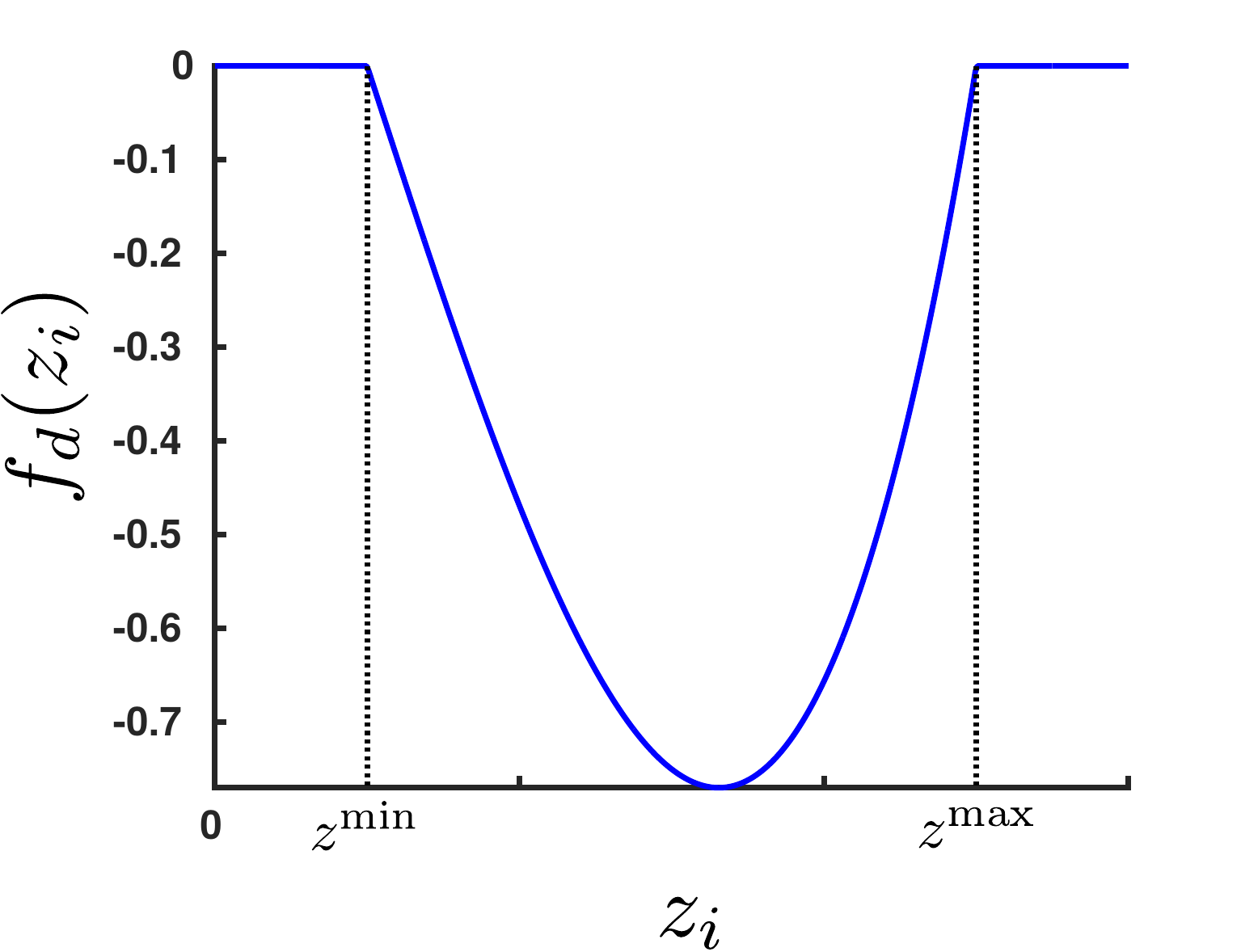}
		\caption{Uniform coverage quality function [Left] and its derivative [Right].}
		\label{fig:quality_plot}
	\end{figure}

	A plot of this function can be seen in Figure \ref{fig:quality_plot} [Left]. This function was chosen so that $f(z^{\min}) = 1$ and $f(z^{\max}) = 0$. In addition, $f(z_i)$ is first order differentiable with respect to $z_i$, or $\frac{\partial f(z_i)}{\partial z_i}$ exists within $C_i^s$, which is a property that will be required when deriving the control law in Section \ref{section:distributed_control_law}.
	
	The derivative $\frac{\partial f(z_i)}{\partial z_i}\colon [z^{\min}, ~z^{\max}] \rightarrow [f_d^{\min}, 0]$ is evaluated as
	\begin{equation*}
	f_d(z_i) \stackrel{\triangle}{=} \frac{\partial f(z_i)}{\partial z_i} = \left \{
	\begin{aligned}
		&~\frac{4 \left( z_i - z^{\min} \right) \left[ \left( z_i - z^{\min} \right)^2 - \left( z^{\max} - z^{\min} \right)^2 \right]}{\left( z^{\max} - z^{\min} \right)^4}, & ~q \in C_i^s\\
		&~0, & ~q \notin C_i^s
	\end{aligned}
	\right.\\
	\end{equation*}
	where $f_d^{\min} = f_d\left( z^{\min} + \frac{\sqrt{3}}{3} \left(z^{\max}-z^{min}\right) \right) = -\frac{8\sqrt{3}}{9\left(z^{\max}-z^{min}\right)}$. A plot of this function can be seen in Figure \ref{fig:quality_plot} [Right].
	
	$f(z_i)$ and $f_d(z_i)$ are 4th and 3rd degree polynomials respectively and as a result continuous functions of $z_i$. It should be noted that any strictly decreasing and differentiable with a continuous derivative function $f(z_i)\colon [z^{\min}, ~z^{\max}] \rightarrow [0, 1]$ can be potentially used.

%%%%%%%%%%%%%%%%%%%%%%%%%%%%%
\section{Sensed space partitioning}
%%%%%%%%%%%%%%%%%%%%%%%%%%%%%%
\label{section:partitioning}
	The assignment of responsibility regions to the nodes is achieved in a manner similar to \cite{Stergiopoulos_ICRA14}, where only the subset of $\Omega$ sensed by the nodes is partitioned. Each node is assigned a cell
	\begin{equation}
	W_i \stackrel{\triangle}{=} \left\{q \in \Omega \colon f(z_i) \geq f(z_j), ~j \neq i \right\}
	\label{partitioning}
	\end{equation}
	with the equality holding true only at the boundary $\partial W_i$, so that the cells $W_i$ comprise a complete tessellation of the sensed region.
	
	Because the coverage quality is uniform, $\partial W_j \cap \partial W_i$ is either an arc of $\partial C_i$ if $z_i < z_j$ or of $\partial C_j$ if $z_i > z_j$. In the case where $z_i = z_j$, $\partial W_j \cap \partial W_i$ is chosen arbitrarily as the line segment defined by the two intersection points of $\partial C_i$ and $\partial C_j$. Hence, the resulting cells consist of circular arcs and line segments.
		
\begin{figure}[htb]
		\centering
		\includegraphics[width=0.49\textwidth]{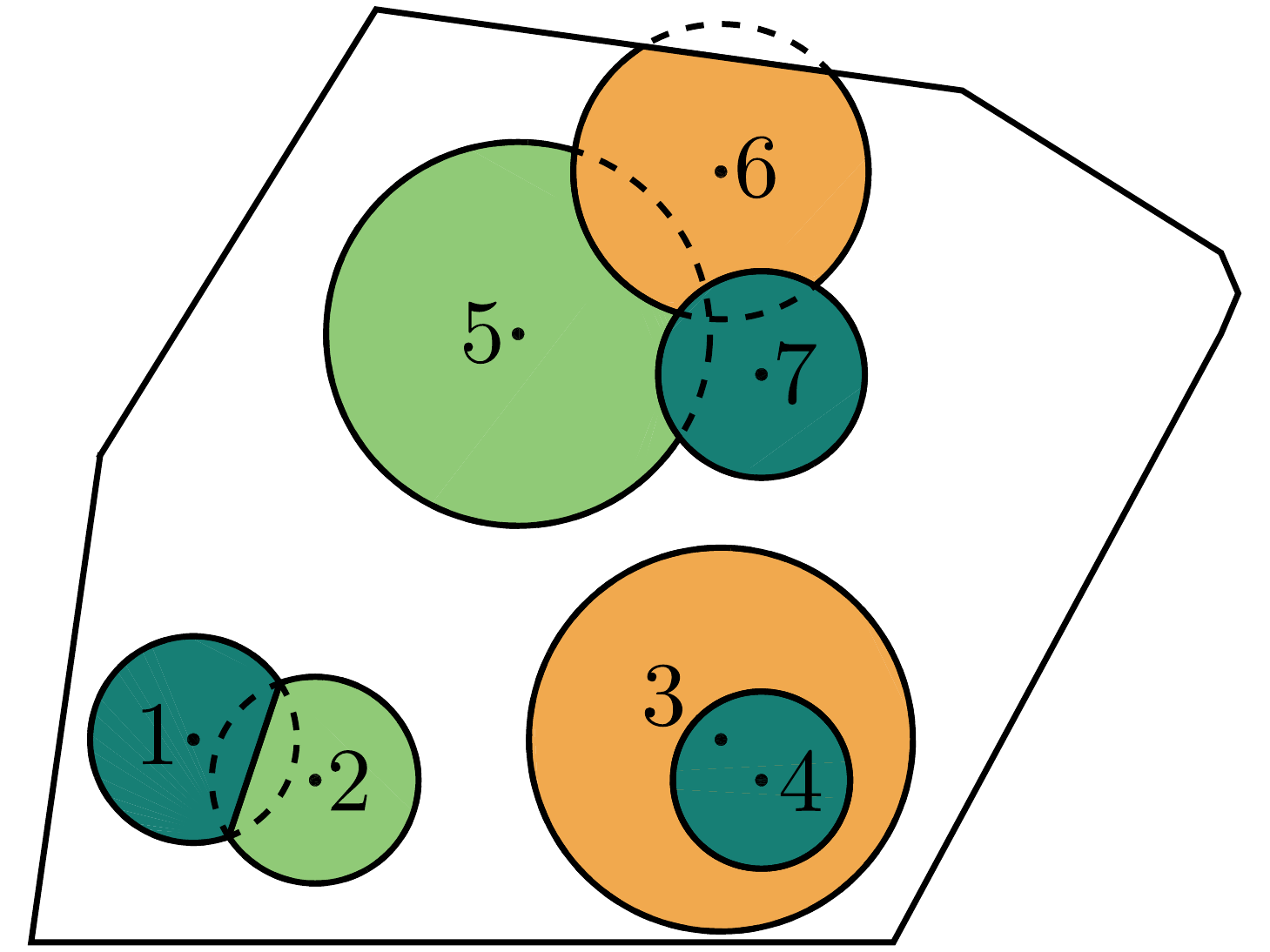}
		\includegraphics[width=0.49\textwidth]{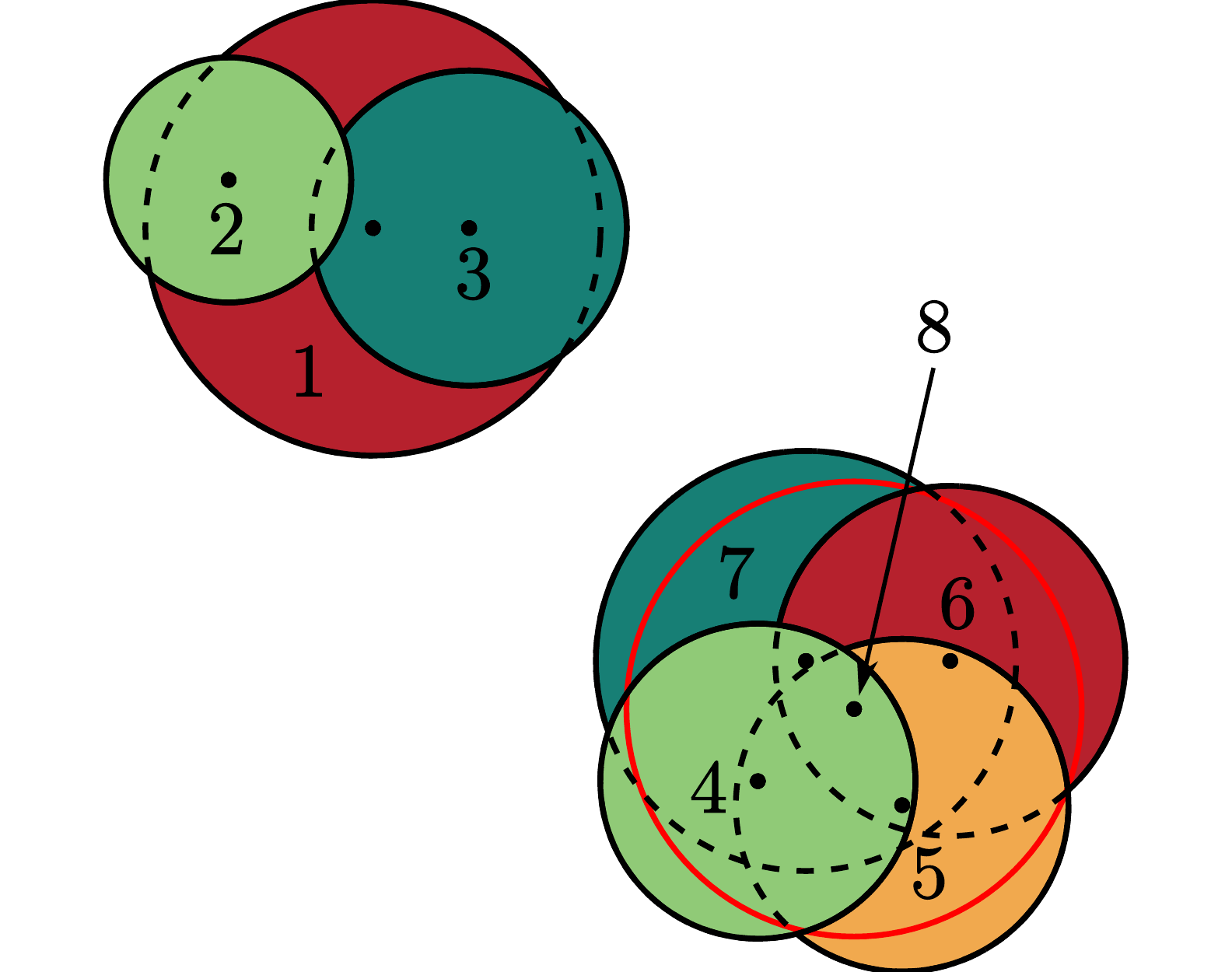}
		\caption{Space partitioning examples.}
		\label{fig:partitioning}
	\end{figure}

	If the sensing disk of a node $i$ is contained within the sensing disk of another node $j$, i.e. $C_i^s \cap C_j^s = C_i^s$, then $W_i = C_i^s$ and $W_j = C_j^s \setminus C_i^s$. An example partitioning with all of the aforementioned cases illustrated can be seen in Figure \ref{fig:partitioning} [Left], where the boundaries of the sensing disks $\partial C_i^s$ are in dashed and the boundaries of the cells $\partial W_i$ in solid black. Nodes $1$ and $2$ are at the same altitude so the arbitrary partitioning scheme is used. The sensing disk of node $3$ contains the sensing disk of node $4$ and nodes $5, 6$ and $7$ illustrate the general case.
	
	By utilizing this partitioning scheme, the network's coverage performance can be written as
	\begin{equation}
	\mathcal{H} = \sum_{i \in I_n} \int_{W_i} f(z_i) ~\phi(q) ~dq.
	\label{criterion}
	\end{equation}
	
	\begin{definition}
	We define the neighbors $N_i$ of node $i$ as
	\begin{equation*}
	N_i \stackrel{\triangle}{=} \left\{ j \neq i \colon C_j^s \cap C_i^s \neq \emptyset \right\}.
	\end{equation*}
	The neighbors of node $i$ are those nodes that sense at least a part of the region that node $i$ senses. It is clear that, due to the partitioning scheme used, only the nodes in $N_i$ need to be considered when creating $W_i$.
	\end{definition}
	\begin{rmk}
	The aforementioned partitioning is a complete tessellation of the sensed region $\bigcup_{i \in I_n} C_i^s$. However it is not a complete tessellation of $\Omega$. The neutral region not assigned by the partitioning scheme is denoted as $\mathcal{O} = \Omega \setminus \bigcup_{i \in I_n} W_i$.
	\end{rmk}
	\begin{rmk}
	The resulting cells $W_i$ are compact but they are not always convex. It is also possible that a cell $W_i$ consists of multiple disjoint regions, such as the cell of node $1$ shown in red in Figure \ref{fig:partitioning} [Right]. In addition it is possible that the cell of a node is empty, such as the cell of node $8$ in Figure \ref{fig:partitioning} [Right]. Its sensing circle $\partial C_8^s$ is shown in a solid red line.
	\end{rmk}
	
%%%%%%%%%%%%%%%%%%%%%%%%%%%%%%
\section{Spatially Distributed Coordination Algorithm}
%%%%%%%%%%%%%%%%%%%%%%%%%%%%%%
\label{section:distributed_control_law}

	Based on the nodes kinodynamics (\ref{kinematics}), their sensing performance (\ref{sensing}) and the coverage criterion (\ref{criterion}), a gradient based control law is designed. The control law utilizes the partitioning (\ref{partitioning}) and result in monotonous increase of the covered area.
	\begin{thm}
	In an MAA visual network consisting of nodes with sensing performance as in (\ref{sensing}), governed by the kinodynamics in (\ref{kinematics})
	and the space partitioning described in Section \ref{section:partitioning}, the control law
	\footnotesize
	\begin{eqnarray}
		%\nonumber
		u_{i,q} &=& \alpha_{i,q} \left[ ~\int\limits_{\partial W_i \cap \partial \mathcal{O}} n_i ~f(z_i) ~dq ~+ \right. %\\
		 \left. \sum\limits_{j \neq i} ~\int\limits_{\partial W_i \cap \partial W_j} \upsilon_i^i ~n_i ~(f(z_i) - f(z_j)) ~dq \right]\\
		%\nonumber
		u_{i,z} &=& \alpha_{i,z} \left[ ~\int\limits_{\partial W_i \cap \partial \mathcal{O}} \tan(a) ~f(z_i) ~dq ~+ f_d(z_i) \int\limits_{W_i}dq ~+ \right. %\\
		 \left. \sum\limits_{j \neq i} ~\int\limits_{\partial W_i \cap \partial W_j} \nu_i^i \cdot n_i ~(f(z_i) - f(z_j)) ~dq \right]
		\label{control_law}
	\end{eqnarray}
	\normalsize
	where $\alpha_{i,q}, \alpha_{i,z}$ are positive constants, $\upsilon_i^i$ and $\nu_i^i$ are the Jacobian matrices of the points $q \in \partial W_i$ with respect to $q_i$ and $z_i$ respectively and $n_i$ the outward pointing normal vector of $W_i$, maximizes the performance criterion (\ref{criterion}) monotonically along the nodes' trajectories, leading in a locally optimal configuration.
	\end{thm}
	
	\begin{pf}
	Initially we evaluate the time derivative of the optimization criterion $\mathcal{H}$
	\begin{equation*}
	\frac{d\mathcal{H}}{dt} = \sum_{i \in I_n} \left[ \frac{\partial\mathcal{H}}{\partial q_i} \dot{q}_i ~+ \frac{\partial\mathcal{H}}{\partial z_i} \dot{z}_i\right]
	 = \sum_{i \in I_n} \left[ \frac{\partial\mathcal{H}}{\partial q_i} u_{i,q} ~+ \frac{\partial\mathcal{H}}{\partial z_i} u_{i,z} \right].
	\end{equation*}.
	
	The usage of a gradient based control law in the form
	\begin{equation*}
	u_{i,q} = \alpha_{i,q} \frac{\partial\mathcal{H}}{\partial q_i}, ~~ u_{i,z} = \alpha_{i,z} \frac{\partial\mathcal{H}}{\partial z_i}
	\end{equation*}
	will result in a monotonous increase of $\mathcal{H}$.
	
	By using the Leibniz integral rule \cite{Flanders_AMM73} we obtain
	\small
	\begin{eqnarray}
	\nonumber
	\frac{\partial\mathcal{H}}{\partial q_i} &=& \sum\limits_{i \in I_n}
	\left[
	~\int\limits_{\partial W_i} \upsilon_i^i ~n_i ~f(z_i) ~dq ~+ \int\limits_{W_i} \frac{\partial f(z_i)}{\partial q_i} ~dq
	\right]\\
	\nonumber
	&=& ~\int\limits_{\partial W_i} \upsilon_i^i ~n_i ~f(z_i) ~dq ~+ \int\limits_{W_i} \frac{\partial f(z_i)}{\partial q_i} ~dq + \sum\limits_{j \neq i} \left[
	~\int\limits_{\partial W_j} \upsilon_j^i ~n_j ~f(z_j) ~dq ~+ \int\limits_{W_j} \frac{\partial f(z_j)}{\partial q_i} ~dq \right]
	\end{eqnarray}
	\normalsize
	where $\upsilon_j^i$ stands for the Jacobian matrix with respect to $q_i$ of the points $q\in \partial W_j$, 
	\begin{equation}
	\upsilon_j^i\left(q\right) \stackrel{\triangle}{=} \frac{\partial q}{\partial q_i},~~q\in \partial W_j,~i,j\in I_n.
	\end{equation}
	
	Since $\frac{\partial f(z_i)}{\partial q_i} = \frac{\partial f(z_j)}{\partial q_i} = 0$ we obtain
	\begin{eqnarray}
	\nonumber
	\frac{\partial\mathcal{H}}{\partial q_i}
	&=& ~\int\limits_{\partial W_i} \upsilon_i^i ~n_i ~f(z_i) ~dq ~+ \sum\limits_{j \neq i}
	~\int\limits_{\partial W_j} \upsilon_j^i ~n_j ~f(z_j) ~dq
	\end{eqnarray}
	whose two terms indicate how a movement of node $i$ affects the boundary of its cell and the boundaries of the cells of other nodes. It is clear that only the cells $W_j$ which have a common boundary with $W_i$ will be affected and only at that common boundary.
	
	The boundary $\partial W_i$ can be decomposed in disjoint sets as
	\begin{equation}
	\hspace{-0.05cm}\partial W_i=
	\left\{\partial W_i \cap \partial \Omega \right\}
	\cup 
	\left\{\partial W_i \cap \partial \mathcal{O} \right\}
	\cup
	\{\bigcup_{j \neq i} \left( \partial W_i \cap \partial W_j \right)\}.
	\label{boundary_decomposition}
	\end{equation}
	These sets represent the parts of $\partial W_i$ that lie on the boundary of $\Omega$, the boundary of the node's sensing region and the parts that are common between the boundary of the cell of node $i$ and those of other nodes. This decomposition can be seen in Figure~\ref{fig:boundary_decomposition} with the sets $\partial W_i \cap \partial \Omega$, $\partial W_i \cap \partial \mathcal{O}$ and $\partial W_i \cap \bigcup_{j \neq i} \partial W_j$ appearing in solid red, green and blue respectively.
	
	\begin{figure}[htb]
		\centering
		\includegraphics[width=0.5\textwidth]{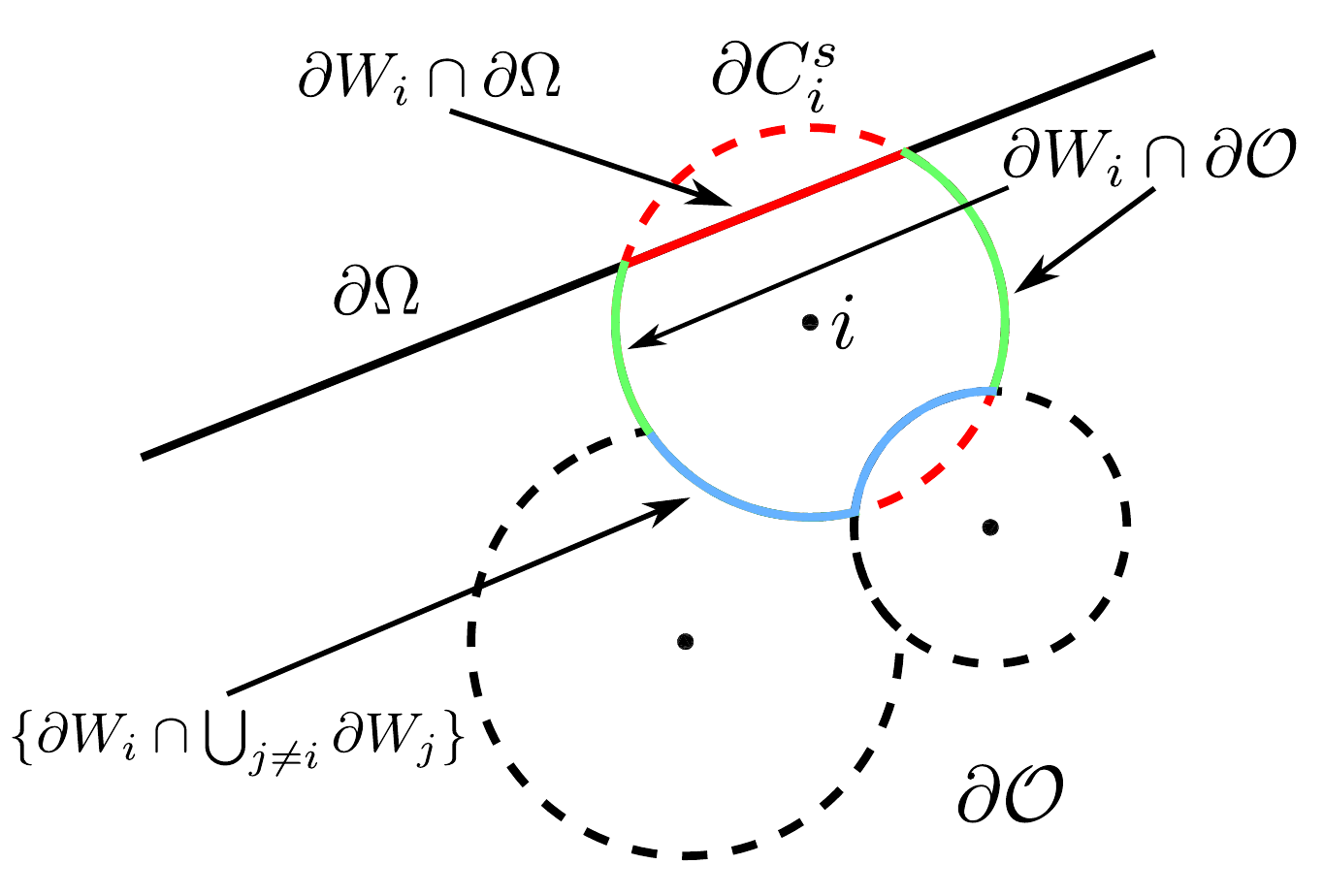}
		\caption{$\partial W_i$-decomposition into disjoint sets}
		\label{fig:boundary_decomposition}
	\end{figure}
	
	At $q \in \partial \Omega$ it holds that $\upsilon_i^i = \textbf{0}_{2 \times 2}$ since we assume the region of interest is static. Additionally, since only the common boundary $\partial W_j \cap \partial W_i$ of node $i$ with any other node $j$ is affected by the movement of node $i$, $\frac{\partial\mathcal{H}}{\partial q_i}$ can be simplified as
	\small
	\begin{eqnarray}
	\nonumber
	\frac{\partial\mathcal{H}}{\partial q_i}
	&=& ~\int\limits_{\partial W_i \cap \partial \mathcal{O}} \upsilon_i^i ~n_i ~f(z_i)  ~dq ~+ \sum\limits_{j \neq i}
	~\int\limits_{\partial W_i \cap \partial W_j} \upsilon_i^i ~n_i ~f(z_i) ~dq ~+ \sum\limits_{j \neq i}
	~\int\limits_{\partial W_j \cap \partial W_i} \upsilon_j^i ~n_j ~f(z_j) ~dq.
	\end{eqnarray}
	\normalsize
	
	The evaluation of $\upsilon_i^i$ can be found in Appendix A. Because the boundary $\partial W_i \cap \partial W_j$ is common among nodes $i$ and $j$, it holds true that $\upsilon_j^i = \upsilon_i^i$ when evaluated over it and that  $n_j = -n_i$. Finally, the sums and the integrals within them can be combined, producing the final form of the planar control law
	\begin{eqnarray}
	\tiny
	\nonumber
	\frac{\partial\mathcal{H}}{\partial q_i}
	&=& ~\int\limits_{\partial W_i \cap \partial \mathcal{O}} ~n_i ~f(z_i)  ~dq ~+ \sum\limits_{j \neq i}
	~\int\limits_{\partial W_j \cap \partial W_i} \upsilon_i^i ~n_i ~\left(f(z_i) - f(z_j)\right) ~dq.
	\normalsize
	\end{eqnarray}
	
	Similarly, by using the same $\partial W_i$ decomposition and defining $\nu_j^i\left(q\right) \stackrel{\triangle}{=} \frac{\partial q}{\partial z_i},~~q\in \partial W_j,~i,j\in I_n$, the altitude control law is
	\small
	\begin{eqnarray}
	\nonumber
	\frac{\partial\mathcal{H}}{\partial z_i}
	&=& ~\int\limits_{\partial W_i \cap \partial \mathcal{O}} ~\nu_i^i \cdot n_i ~f(z_i)  ~dq ~+ \int\limits_{W_i} \frac{\partial f(z_i)}{\partial z_i} ~dq ~+ \sum\limits_{j \neq i}
	~\int\limits_{\partial W_j \cap \partial W_i} \nu_i^i \cdot n_i ~\left(f(z_i) - f(z_j)\right) ~dq
	\end{eqnarray}
	\normalsize
	where the evaluation of $\nu_i^i(q) \cdot n_i$ on $\partial W_i \cap \partial \mathcal{O}$ and $\partial W_j \cap \partial W_i$ can also be found in Appendix A. Because $\frac{\partial f(z_i)}{\partial z_i}$ is constant over $W_i$ and using the expression for $\nu_i^i(q) \cdot n_i$ from Appendix A, the control law can be further simplified into
	\small
	\begin{eqnarray}
	\nonumber
	\frac{\partial\mathcal{H}}{\partial z_i}
	&=& ~\int\limits_{\partial W_i \cap \partial \mathcal{O}} \tan(a) ~f(z_i)  ~dq ~+ f_d(z_i) \int\limits_{W_i}dq ~+ \sum\limits_{j \neq i}
	~\int\limits_{\partial W_j \cap \partial W_i} \nu_i^i \cdot n_i ~\left(f(z_i) - f(z_j)\right) ~dq.
	\end{eqnarray}
	\normalsize
	
	\end{pf}
	
	\begin{rmk}
	The cell $W_i$ of node $i$ is affected only by its neighbors $N_i$ thus resulting in a distributed control law. The discovery of the neighbors $N_i$ depends on their coordinates $X_j,~j \in N_{i}$ and does not correspond to the classical 2D-Delaunay neighbor search. The computation of the $N_i$ set demands node $i$ to be able to communicate with all nodes within a sphere centered around $X_i$ and radius $r_{i}^{c}$
	\small
	\begin{eqnarray}
	\nonumber
	r_{i}^{c}=\max\left\{2~z_i~\tan a,~\left(z_i+z^{\min}\right)^2 \tan^2 a + \left(z_i-z^{\min}\right)^2,~\left(z_i+z^{\max}\right)^2 \tan^2 a + \left(z_i-z^{\max}\right)^2\right\}.
	\end{eqnarray}
	\normalsize
	Figure~\ref{fig:neighbor_set} highlights the case where nodes $2$, $3$ and $4$ are at $z^{\min}$, $z_1$ and $z^{\max}$ respectively. These are the worst case scenario neighbors of node $1$ , the farthest of which dictates the communication range $r_1^c$. 

	\begin{figure}[htb]
		\centering
		\includegraphics[width=0.5\textwidth]{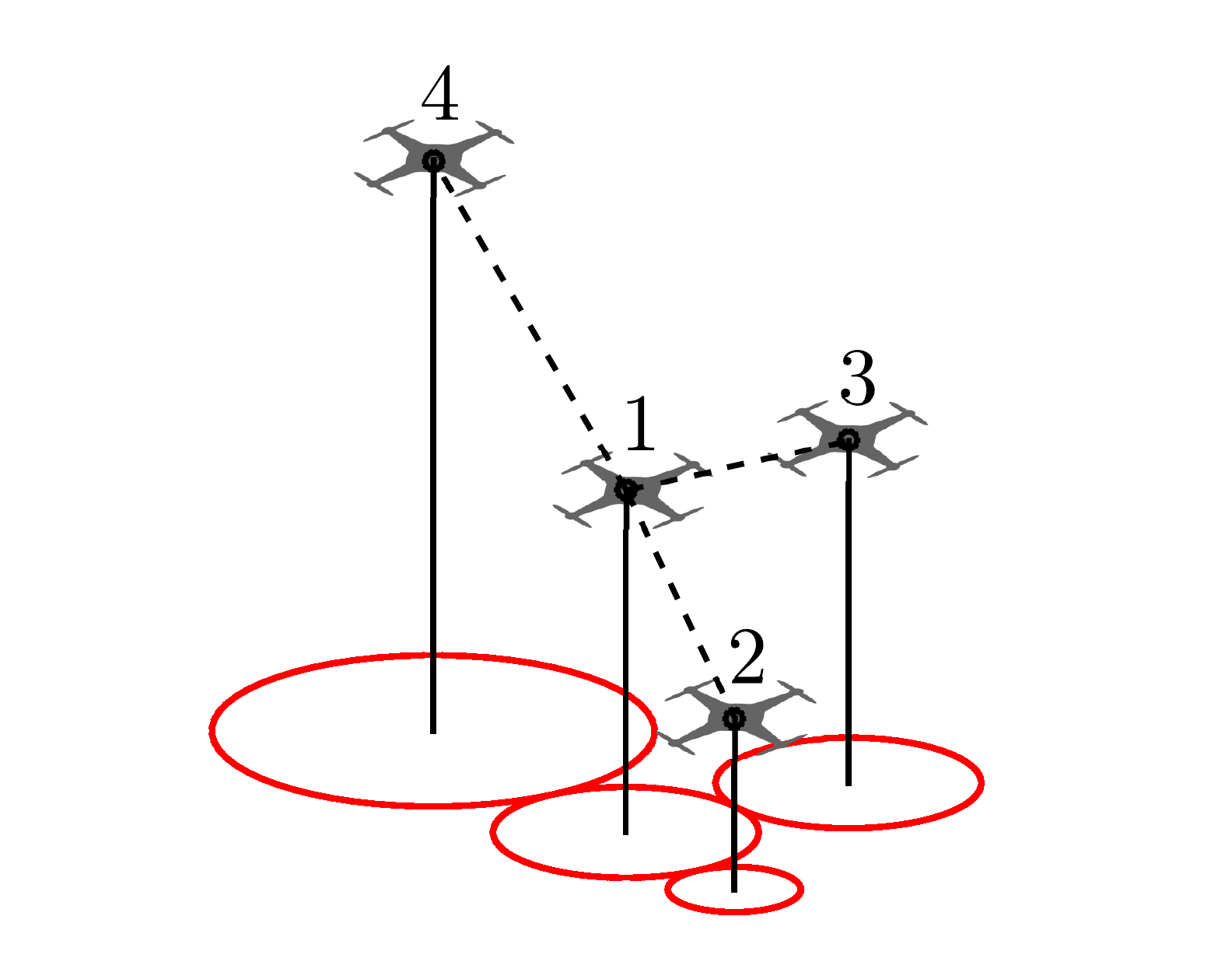}
		\caption{$N_i$ neighbor set}
		\label{fig:neighbor_set}
	\end{figure}
	\end{rmk}
	
	\begin{rmk}
	\label{remark:zmax}
	When $z_i = z^{\max}$, both the planar and altitude control laws are zero because $f(z_i) = 0$. This results in the MAA being unable to move any further in the future and additionally its contribution to the coverage-quality objective being zero. However this degenerate case is of little concern, as shown in Sections \ref{section:stable_altitude} and \ref{section:degenerate_cases}.
	\end{rmk}
	
	\begin{rmk}
	The control law essentially maximizes the volume contained by the union of all the cylinders defined by $f(z_i), ~i \in I_n$, under the constraints imposed by the network and area of interest.
	\end{rmk}

	\section{MAA Altitude Stability}
	\label{section:stability}
	In this section we examine the stability of the nodes' altitude $z_i$ and show that it always remains in the interval $[z^{\min}, ~z^{\max}]$. The system under examination is
	\begin{equation*}
	\dot{z}_i = u_{i,z}, ~u_{i,z}\in \mathbb{R}.
	\end{equation*}
	We will first find and characterize its equilibrium points for the case of a single node and then generalize to the case of multiple nodes.

	\subsection{Optimal altitude for a single MAA}
	\label{sec:optimal_altitude}
	It is useful to define an optimal altitude $z^{opt}$ as the altitude a node would reach if: 1) it had no neighbors $(N_i = \varnothing)$, and 2) its whole cell was inside the region of interest $(\Omega \cap W_i = W_i)$. When the aforementioned requirements are met it holds true that $W_i =  C_i^s$. This optimal altitude is the stable equilibrium point of the system
	\begin{equation*}
	\dot{z}_i = u_{i,z}^{opt}, ~u_{i,z}^{opt}\in \mathbb{R}
	\end{equation*}
	where
	\begin{eqnarray}
	\nonumber
	u_{i,z}^{opt} = \int\limits_{\partial C_i^s} \tan(a) ~f(z_i) ~dq ~+ f_d(z_i) \int\limits_{ C_i^s} ~dq = 2 \pi ~\tan^2(a) ~z_i f(z_i) ~+ \pi ~\tan^2(a) ~z_i^2 ~f_d(z_i)
	\end{eqnarray}
	
	Its value and stability are examined in the following section. This altitude is constant and depends solely on the network's parameters $z^{\min}$ and $z^{\max}$. Had we allowed the nodes to have different minimum and maximum altitudes, each node would have a different constant optimal altitude $z_i^{opt}$.
	
	Additionally, let us denote the sensing region of a node $i$ at $z^{opt}$ as $C_{i,opt}^s\left([x_i~y_i~z_i^{opt}]^T,a\right)$ and $\mathcal{H}_{opt}$ the value of the criterion when all nodes are located at $z^{opt}$. 
	
	If $\Omega = \mathbb{R}^2$ and because the planar control law $u_{i,q}$ results in the repulsion of the nodes, the network will reach a state in which no node will have neighbors and all nodes will be at $z^{opt}$. In that state, the coverage-quality criterion (\ref{criterion}) will have attained its maximum possible value $\mathcal{H}_{opt}$ for that particular network configuration and coverage quality function $f$. This network configuration will be globally optimal. 
	
	When $\Omega$ is a convex compact subset of $\mathbb{R}^2$, it is possible for the network to reach a state where all the nodes are at $z^{opt}$ only if $n$ $C_{i,opt}^s$ disks can be packed inside $\Omega$. This state will be globally optimal. If that is not the case, the nodes will converge at some altitude other than $z^{opt}$ and in general different among nodes. It should be noted that although the nodes do not reach $z^{opt}$, the network configuration is locally optimal.

	\subsection{Optimal altitude stability}
	\label{sec:optimal_altitude_stability}
	We will now evaluate $z^{opt}$ and its stability properties. The system under examination is
	\begin{eqnarray}
	\nonumber
	\dot{z}_i &=& u_{i,z}^{opt}.
	\end{eqnarray}
	In Appendix B it is shown that out of the five equilibrium points of this system, only two reside in the interval $[z^{\min}, ~z^{\max}]$. Those are
	\begin{eqnarray}
	\label{eq_points}
	\nonumber
	z_2^{eq} &=& z^{\max} \\
	\nonumber
	z_5^{eq} &=& \frac{2}{3}~z^{\min} + \frac{1}{3} \sqrt{Q}
	\end{eqnarray}
	where 
\begin{equation}
Q  = 3~{z^{\max}}^2 - 6~z^{\max}~z^{\min} + 4~{z^{\min}}^2 = ~3\left(z^{\max}-z^{\min}\right)^2 + {z^{\min}}^2 = P+{z^{\min}}^2 > 0.
\label{Q_P_definition}
\end{equation}
	
\begin{figure}[htb]
		\centering
		\includegraphics[width=0.5\textwidth]{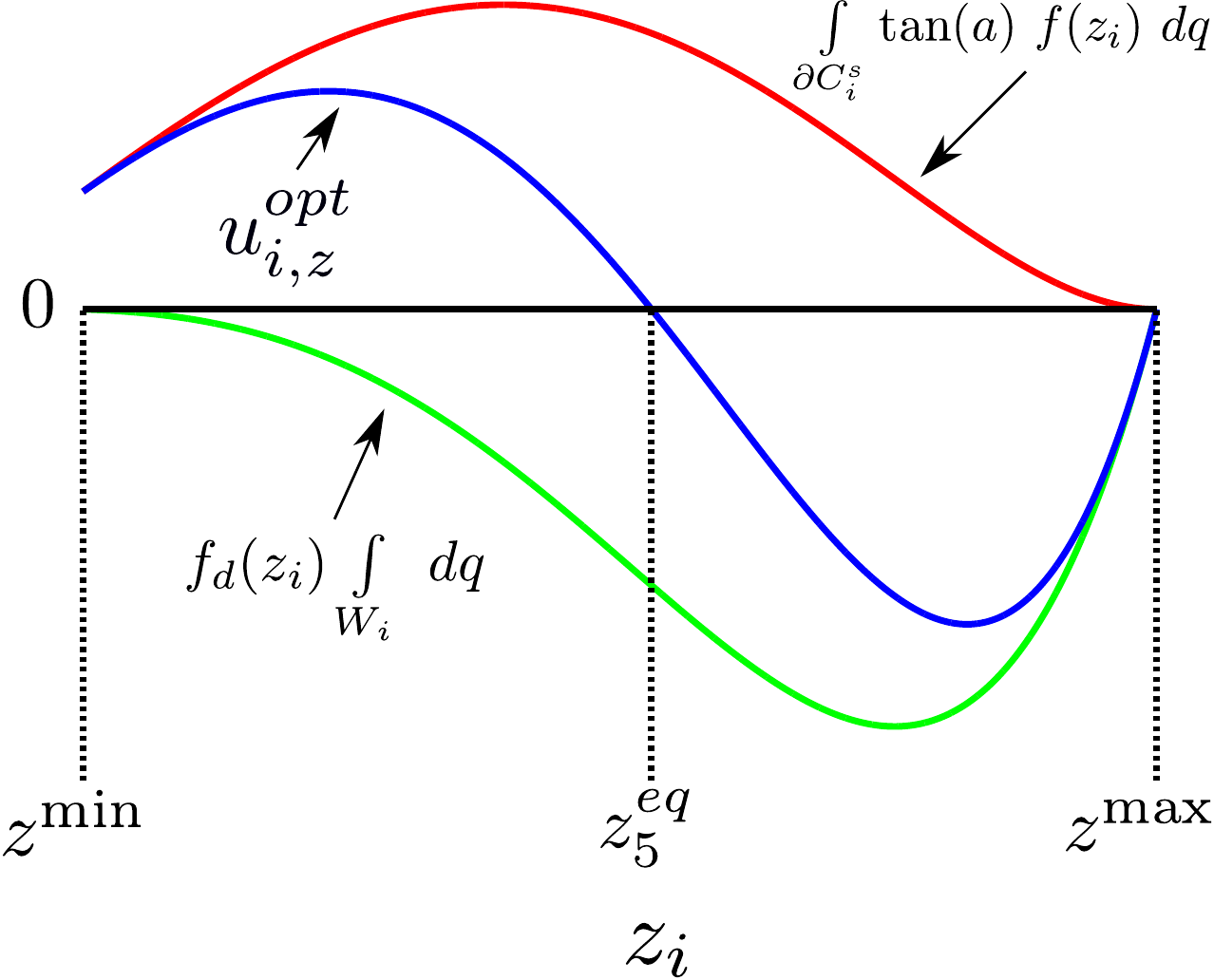}
		\caption{Plot of $u_{i,z}^{opt}$ and its terms over $W_i$ and $\partial W_i$ with respect to $z_i$.}
		\label{fig:u_optimal}
	\end{figure}

	Because the system is scalar, in order to evaluate the stability of those two equilibrium points, it is sufficient to consider the sign of $u_{i,z}^{opt}$ in the interval $[z^{\min}, ~z^{\max}]$. Since $u_{i,z}^{opt}$ is continuous in $[z^{\min}, ~z^{\max}]$, its sign will be constant between consecutive roots of $u_{i,z}^{opt} = 0$. It is shown in Appendix C that 
	\begin{eqnarray}
	\nonumber
	u_{i,z}^{opt} &>& 0, ~~~\forall z_i \in \left[z^{\min}, z_5^{eq}\right) \\
	\nonumber
	u_{i,z}^{opt} &<& 0, ~~~\forall z_i \in \left(z_5^{eq}, z^{\max}\right).
	\end{eqnarray}
	This can also be seen in Figure \ref{fig:u_optimal} where $u_{i,z}^{opt}(z_i)$ is shown in blue, the integral over $W_i$ in green and the integral over $\partial W_i$ in red.
	
	It can now be shown that the equilibrium point $z^{\max}$ is unstable because a small negative disturbance $dz$ will result in $u_{i,z}^{opt} < 0$, thus leading the node to a lower altitude and away from $z^{\max}$.
	
	Similarly, the equilibrium point $z_5^{eq}$ is asymptotically stable. This is because a small negative disturbance $dz$ will result in $u_{i,z}^{opt} > 0$, thus leading the node to a higher altitude and closer to $z_5^{eq}$. Conversely, a small positive disturbance $dz$ will result in $u_{i,z}^{opt} < 0$, thus leading the node to a lower altitude and again closer to $z_5^{eq}$.
	
	To conclude, when a node has no neighbors and its whole cell is inside $\Omega$, the only stable equilibrium point is $z^{opt} = z_5^{eq}$ which has a domain of attraction $[z^{\min}, ~z^{\max})$.

	\subsection{Stable altitude for a team of MAAs}
	\label{section:stable_altitude}
	In the general case, each node will move towards an altitude which is an equilibrium point of the system
	\begin{equation}
	\label{eq:system_altitude}
	\dot{z}_i = u_{i,z}, ~u_{i,z}\in \mathbb{R}
	\end{equation}
	where
	\small
	\begin{equation}
	\label{eq:system_altitude_law}
	u_{i,z} = \int\limits_{\partial W_i \cap \partial \mathcal{O}} \tan(a) ~f(z_i) ~dq ~+ f_d(z_i) \int\limits_{W_i} dq ~+ \sum\limits_{j \neq i} ~\int\limits_{\partial W_i \cap \partial W_j} \tan(a) ~(f(z_i) - f(z_j)) ~dq
	\end{equation}
	\normalsize
	We call this the stable altitude $z_i^{stb}$. The stable altitude is not common among nodes as it depends on one's neighbors $N_i$ and is not constant over time since the neighbors change over time.
	
	We will attempt to generalize the proof of Section \ref{sec:optimal_altitude_stability} in the case of a node with neighbors, which is the general case. The system under examination is derived from equations (\ref{eq:system_altitude}) and (\ref{eq:system_altitude_law}).	The integrals over $\partial W_i$ are non--negative whereas the integral over $W_i$ is non--positive. The integrals over $\partial W_i$ of a node with neighbors will always be smaller than the same integral of a node without neighbors. This is because the neighbors will either remove some arcs of $W_i$ from the integral or reduce their influence due to the term $f(z_i) - f(z_j)$. Similarly, the absolute value of the integral over $W_i$ of a node with neighbors will not be greater than the same integral of a node without neighbors. This is due to the area of $W_i$ possibly being reduced because part of $C_i^s$ has been assigned to neighbors with higher coverage quality. Thus we conclude that $z_i^{stb}$ will attain its minimum value when the integrals over $\partial W_i$ are zero and its maximum value when the integral over $W_i$ is zero.
	
	When the integrals over $\partial W_i$ are both zero, the control law $u_{i,z}$ has a negative value. This will lead to a reduction of the node's altitude and in time the node will reach $z_i^{stb} = z^{\min}$, provided the integrals over $\partial W_i$ remain zero. Once the node reaches $z^{\min}$ its altitude control law will be $0$ until the integral over $\partial W_i$ stops being zero. The planar control law $u_{i,q}$ however is unaffected, so the node's performance in the future is not affected. This situation may arise in a node with several neighboring nodes at lower altitude that result in $\partial C_i^s \cap \partial W_i = \emptyset$.
	
	When the integral over $W_i$ is zero, the control law $u_{i,z}$ has a positive value. This will lead to an increase of the node's altitude and in time the node will reach $z_i^{stb} = z^{\max}$ and as shown in Remark \ref{remark:zmax} the node will be immobilized from this time onwards. However this situation will not arise in practice as explained in Section \ref{section:degenerate_cases}.
	
	When the integral over $W_i$ and at least one of the integrals over $\partial W_i$ are non-zero, then ${z_i^{stb} \in \left(z^{\min}, z^{\max}\right)}$.
	
	The stability of $z_i^{stb}$ is shown similarly to the stability of $z^{opt}$, by using the sign of $u_{i,z}$.

	\subsection{Degenerate cases}
	\label{section:degenerate_cases}
	It is possible due to the nodes' initial positions that the sensing disk of some node $i$ is completely contained within the sensing disk of another node $j$, i.e. $C_i^s \cap C_j^s = C_i^s$. In such a case, it is not guaranteed that the control law will result in separation of the nodes' sensing regions and thus it is possible that the nodes do not reach $z^{opt}$. Instead, node $j$ may converge to a higher altitude and node $i$ to a lower altitude than $z^{opt}$, while their projections on the ground $q_i$ and $q_j$ remain stationary. Because the region covered by node $i$ is also covered by node $j$, the network's performance is impacted negatively. Since this degenerate case may only arise at the network's initial state, care must be taken to avoid it during the agents' deployment. Such a degenerate case is shown in Figure \ref{fig:partitioning} [Left] where the sensing disk of node $4$ is completely contained within that of node $3$.
	
	Another case of interest is when some node $i$ is not assigned a region of responsibility, i.e. ${W_i  = \emptyset}$. This is due to the existence of other nodes at lower altitude that cover all of $C_i^s$ with better quality than node $i$. This is the case with node $8$ in Figure \ref{fig:partitioning} [Right]. This situation is resolved since the nodes at lower altitude will move away from node $i$ and once node $i$ has been assigned a region of responsibility it will also move. It should be noted that the coverage objective $\mathcal{H}$ remains continuous even when node $i$ changes from being assigned no region of responsibility to being assigned some region of responsibility.
	
	In order for a node to reach $z^{\max}$, as explained in Section \ref{section:stable_altitude}, the integral over $W_i$ of its altitude control law $u_{i,z}$ must be zero, that is its cell must consist of just a closed curve without its interior. In order to have $W_i = \partial W_i$, a second node $j$ must be directly below node $i$ at an infinitesimal distance. However just as node $i$ starts moving upwards the integral over $W_i$ will stop being zero thus changing the stable altitude to some value $z_i^{stb} < z^{\max}$. In other words, in order for a node to reach $z^{\max}$, the configuration described must happen at an altitude infinitesimally smaller than $z^{\max}$. So in practice, if all nodes are deployed initially at an altitude smaller than $z^{\max}$, no node will reach $z^{\max}$ in the future.

%%%%%%%%%%%%%%%%%%%%%%%%%%%%%%%%%%%%%%%%%
\section{Simulation Studies}
\label{section:simulation_studies}
	
	Simulation results of the proposed control law using the uniform coverage quality function $f$ are presented in this section. The region of interest $\Omega$ is the same as the one used in \cite{Stergiopoulos_IETCTA10} for consistency. All nodes are identical with a half sensing cone angle $a = 20^{\circ}$ and $z_i \in [0.5, ~2.5], ~\forall i \in I_n$. The boundaries of the nodes' cells are shown in solid black and the boundaries of their sensing disks in dashed red lines.
	
	\begin{rmk}
	It is possible to observe jittering on the cells of some nodes $i$ and $j$. This can happen when $z_i = z_j$ and the arbitrary boundary $\partial W_i \cap \partial W_j$ is used. Once the altitude of one of the nodes changes slightly, the boundary between the cells will change instantaneously from a line segment to a circular arc. The coverage-quality objective $\mathcal{H}$ however will present no discontinuity when this happens.
	\end{rmk}
	
	\subsection{Case Study I}
\begin{figure}[htbp]
		\centering
		\includegraphics[width=0.49\textwidth]{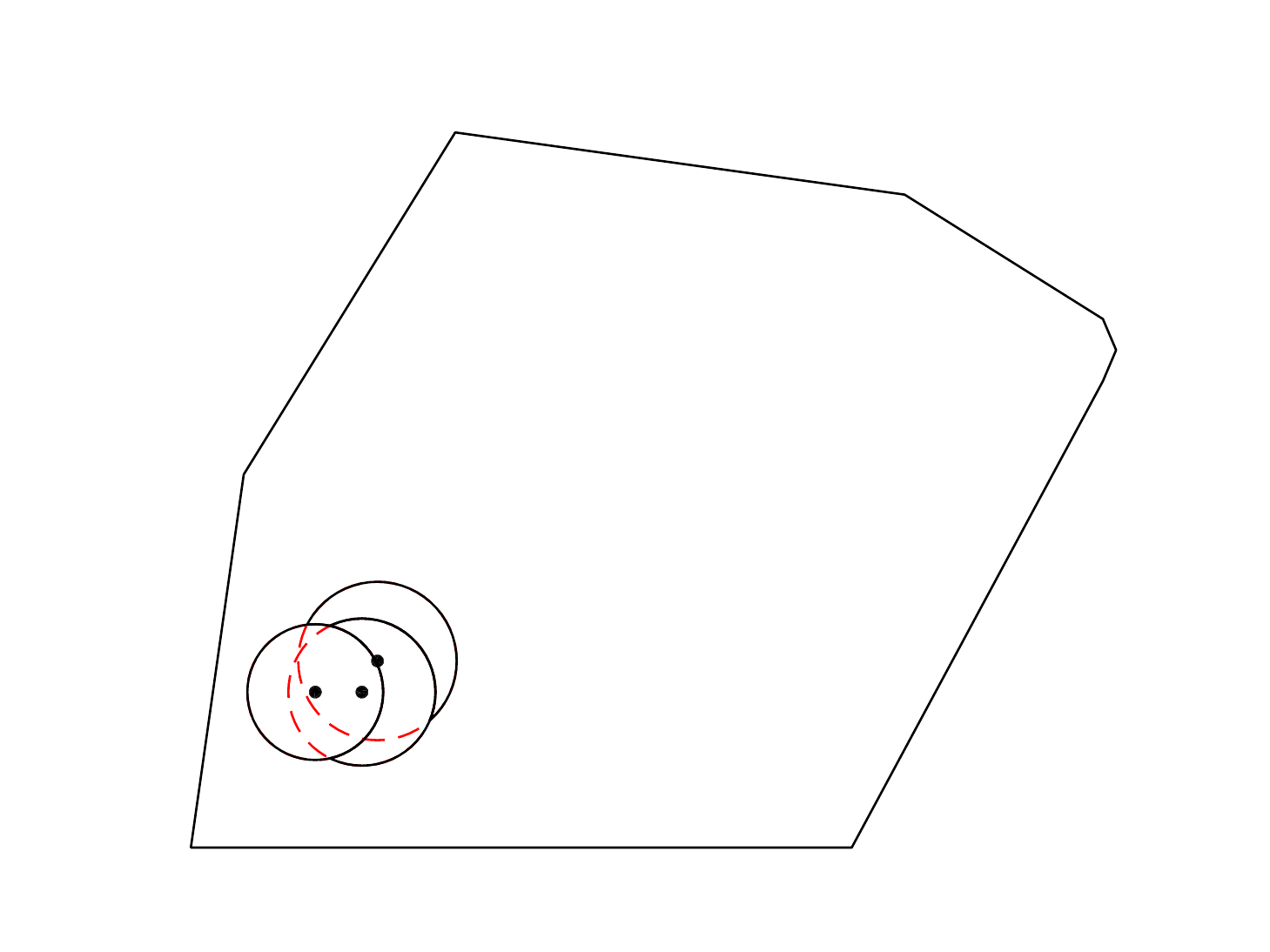}\hspace{0.01cm}
		\includegraphics[width=0.49\textwidth]{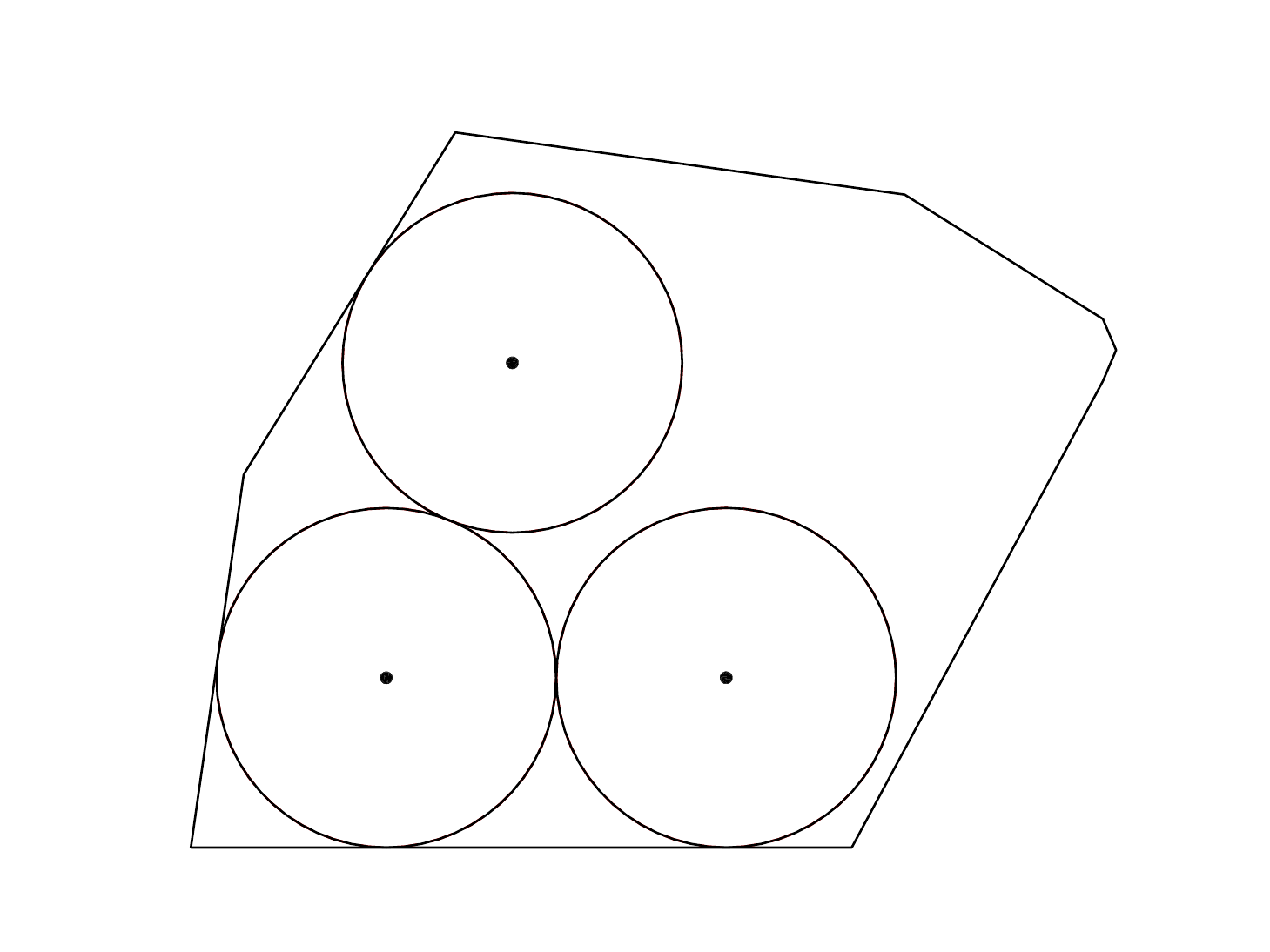}
		\caption{Initial [Left] and final [Right] network configuration and space partitioning.}
		\label{fig:uniform_3_nodes_2D}
	\end{figure}

\begin{figure}[htbp]
		\centering
		\includegraphics[width=0.49\textwidth]{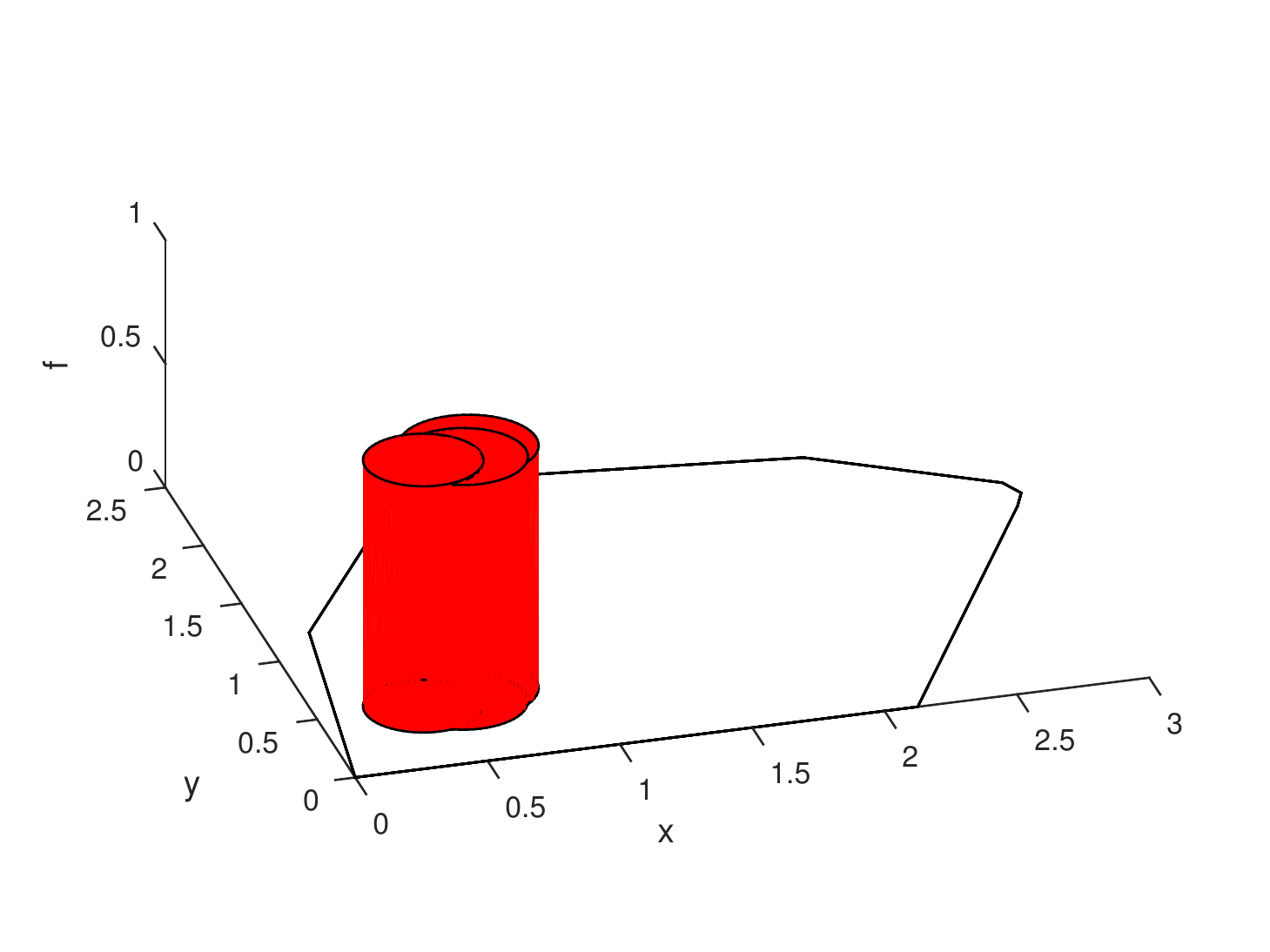}\hspace{0.01cm}
		\includegraphics[width=0.49\textwidth]{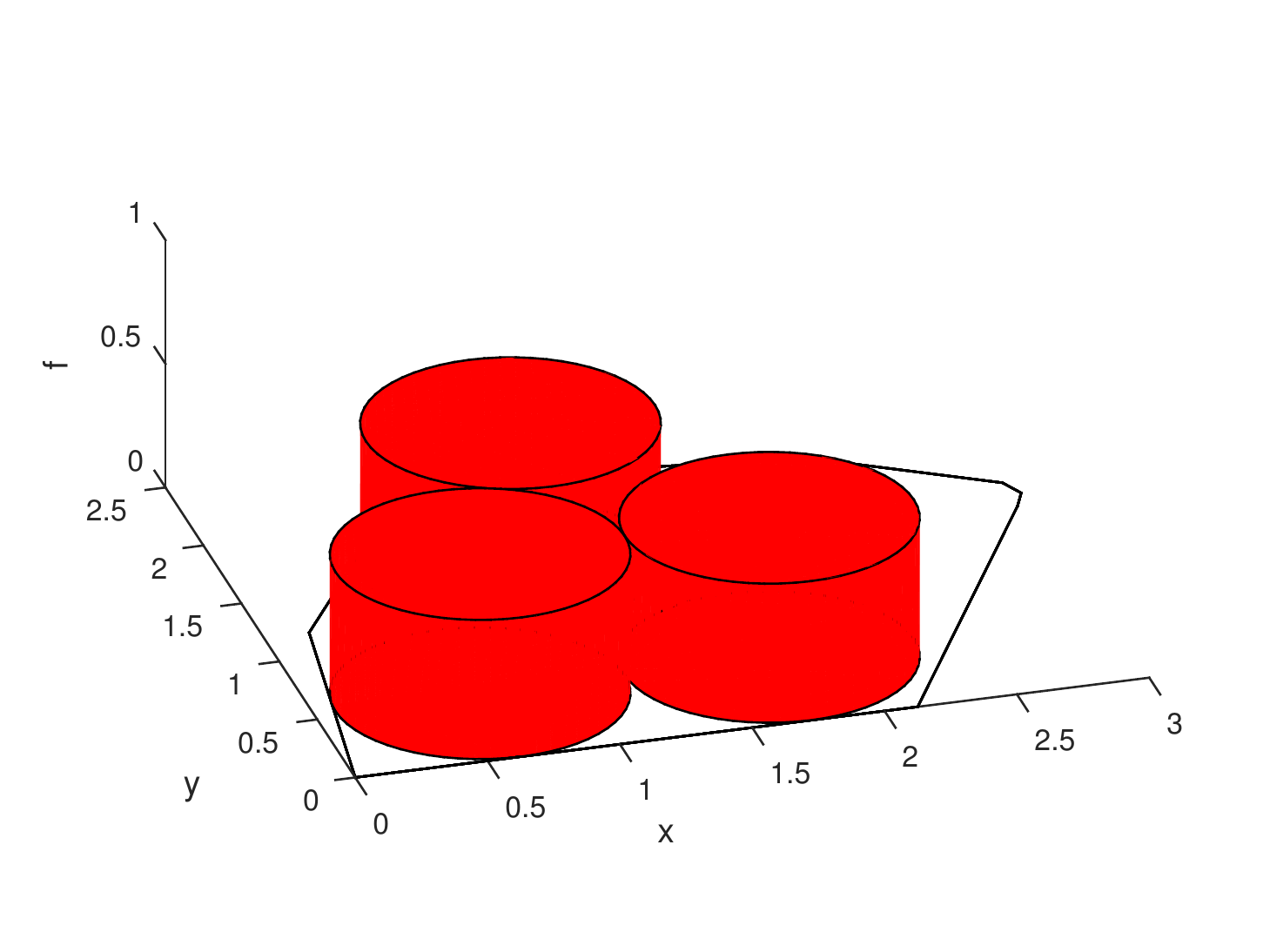}
		\caption{Initial [Left] and final [Right] coverage quality.}
		\label{fig:uniform_3_nodes_3D}
	\end{figure}

\begin{figure}[htbp]
		\centering
		\includegraphics[width=0.8\textwidth]{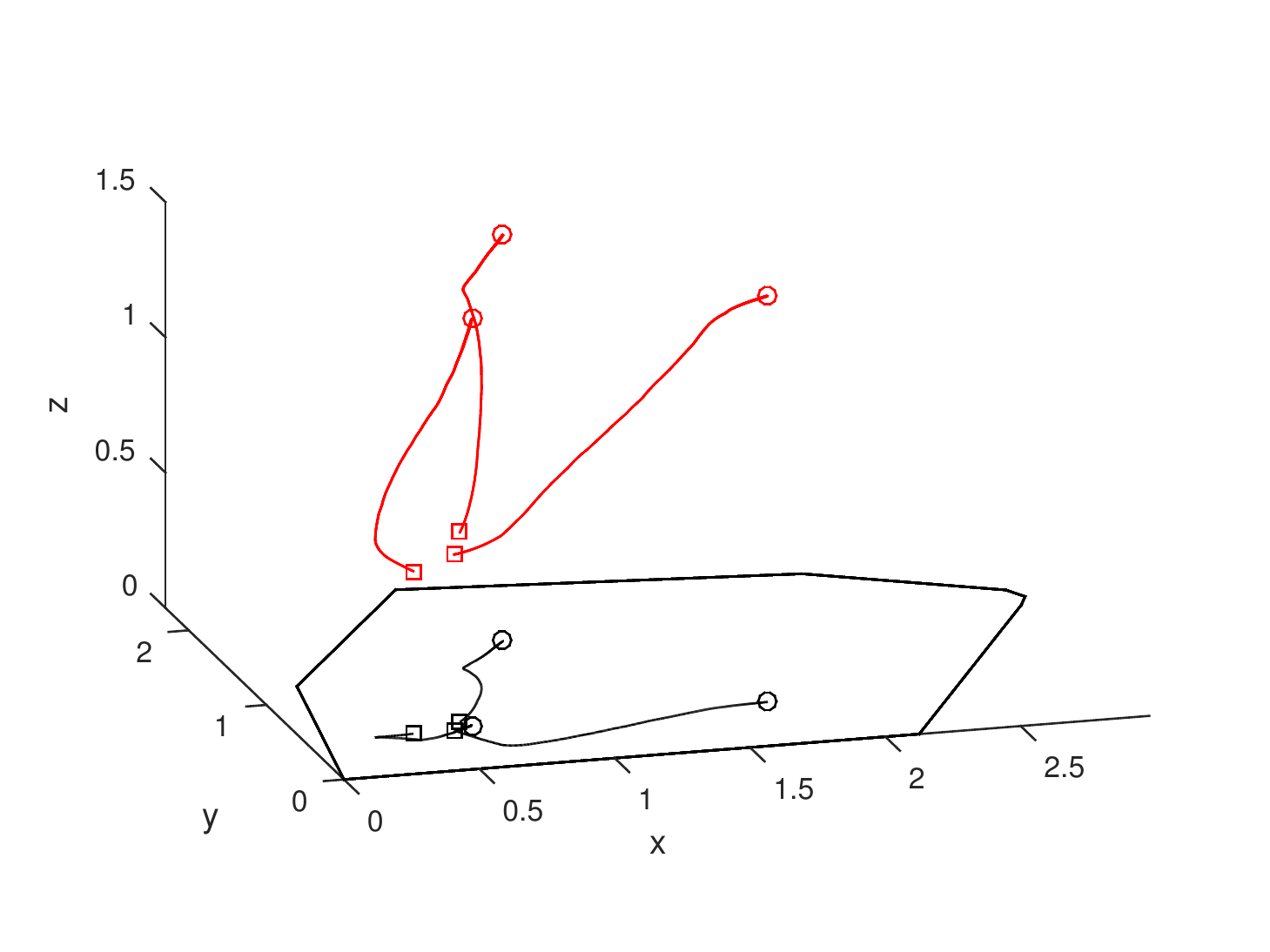}
		\caption{Node trajectories (blue) and their projections on the sensed region (black).}
		\label{fig:uniform_3_nodes_traj}
	\end{figure}

\begin{figure}[htbp]
		\centering
		\includegraphics[width=0.49\textwidth]{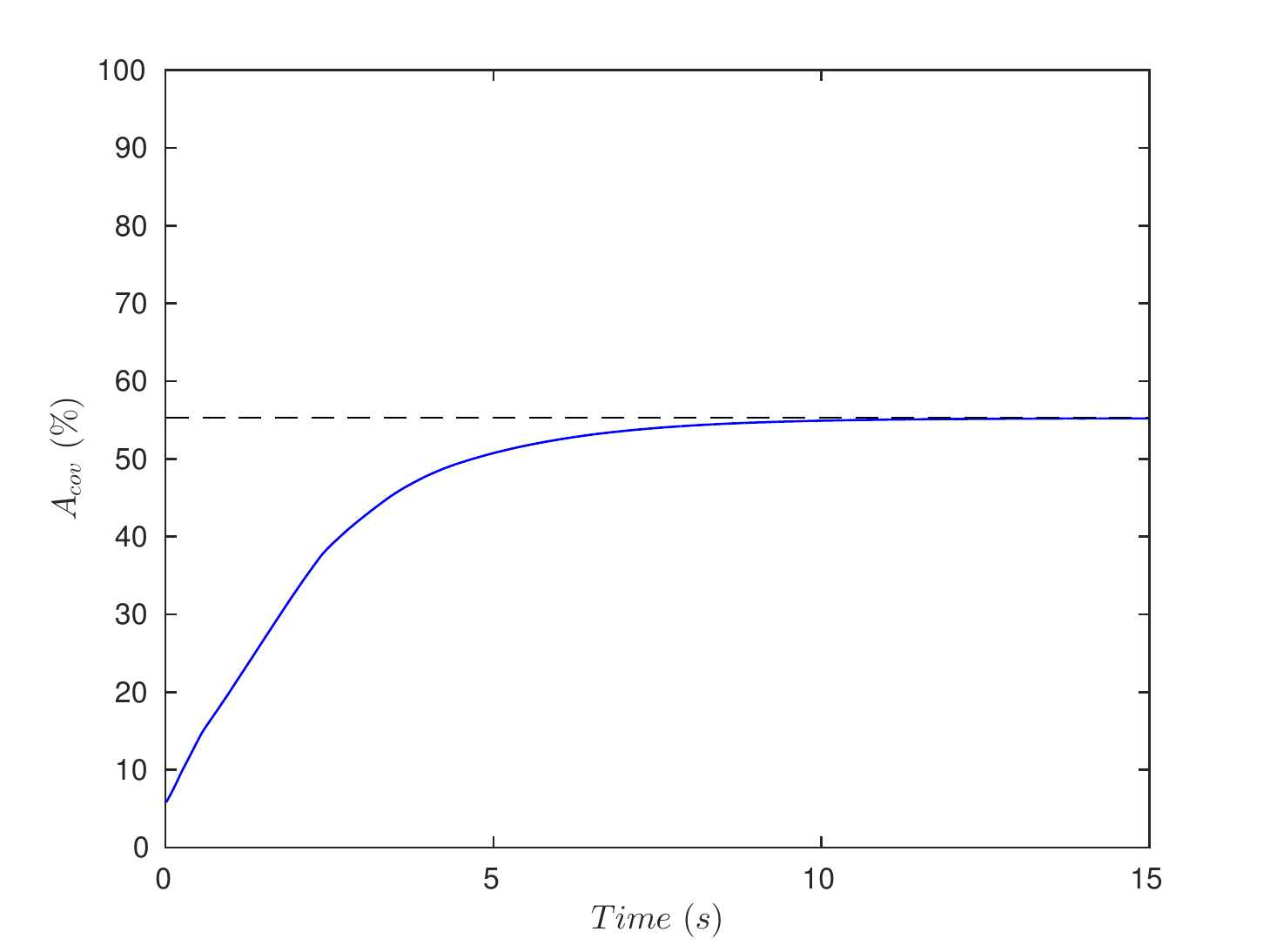}\hspace{0.01cm}
		\includegraphics[width=0.49\textwidth]{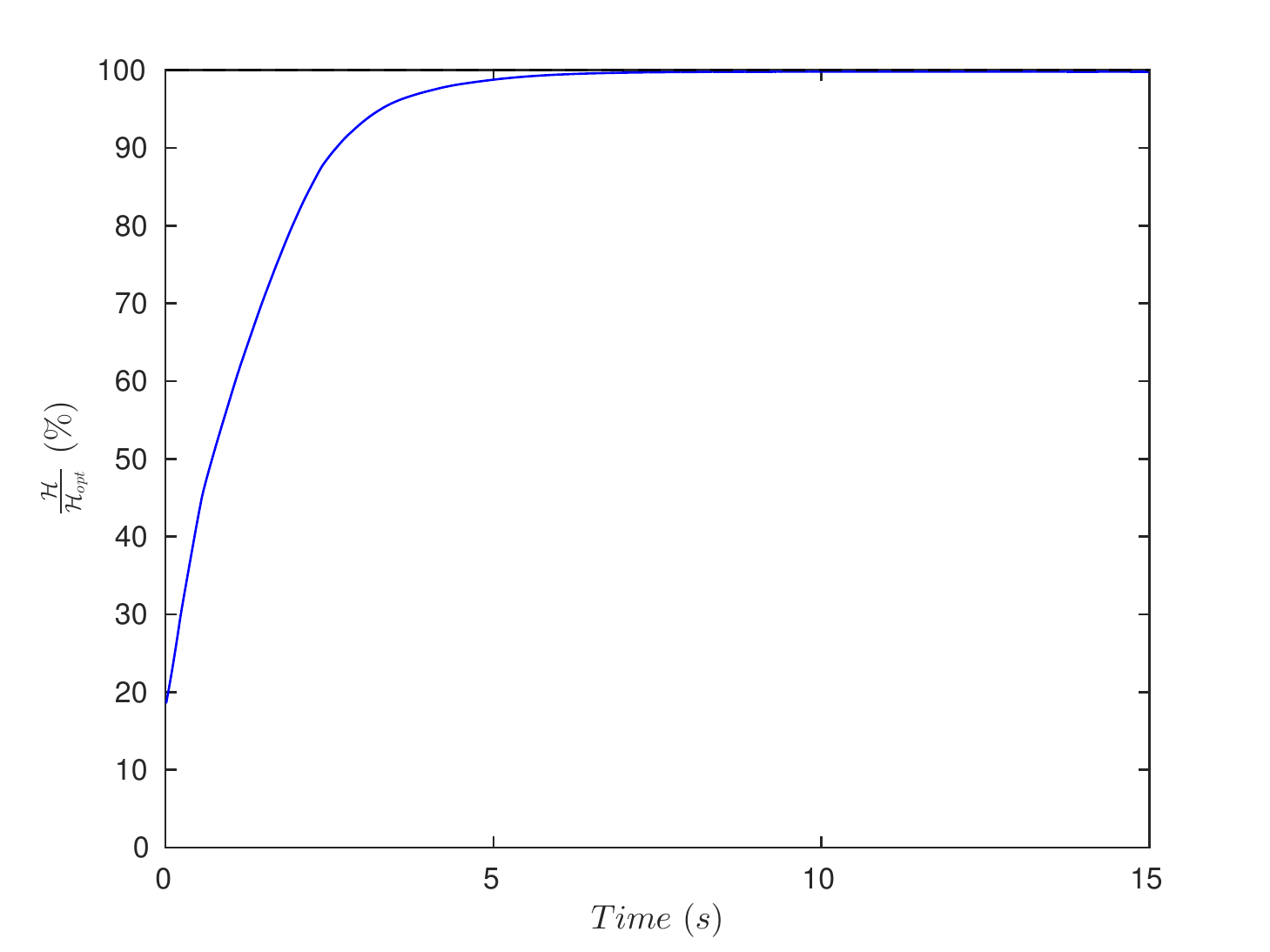}
		\caption{$\frac{\mathcal{A}\left( \bigcup_{i \in I_n} C_i^s \right)}{\mathcal{A}(\Omega)}$ [Left] and $\frac{\mathcal{H}}{\mathcal{H}_{opt}}$ [Right].}
		\label{fig:uniform_3_nodes_area}
	\end{figure}

	In this simulation three nodes start grouped as seen in Figure (\ref{fig:uniform_3_nodes_2D}) [Left]. Since the region of interest is large enough for three optimal disks $C_{i,opt}^s$ to fit inside, all the nodes converge at the optimal altitude $z^{opt}$. As it can be seen in Figure~\ref{fig:uniform_3_nodes_area}, the area covered by the network is equal to $\mathcal{A}\left( \bigcup_{i \in I_n} C_{i,opt}^s \right)$ and the coverage-quality criterion has reached $\mathcal{H}_{opt}$ = $\mathcal{A}\left( \bigcup_{i \in I_n} C_{i,opt}^s \right)$. However since all nodes converged at $z^{opt}$, the addition of more nodes will result in significantly better performance coverage and quality wise, as is clear from Figure \ref{fig:uniform_3_nodes_2D} [Right] and Figure \ref{fig:uniform_3_nodes_area} [Left]. Figure \ref{fig:uniform_3_nodes_3D} shows a graphical representation of the coverage quality at the initial and final stages of the simulation. It is essentially a plot of all $f(z_i)$ inside the region of interest. The volume of the cylinders in Figure \ref{fig:uniform_3_nodes_3D} [Right] is the maximum possible. The trajectories of the  MAAs in $\mathbb{R}^3$ can be seen in Figure \ref{fig:uniform_3_nodes_traj} in red and their projections on the region of interest in black. The initial positions of the  MAAs are marked by squares and their final positions by circles.

	% % % % % % % % % % % % FIGURES 3 NODES % % % % % % % % % % % %
	
	\subsection{Case Study II}
\begin{figure}[htbp]
			\centering
			\includegraphics[width=0.49\textwidth]{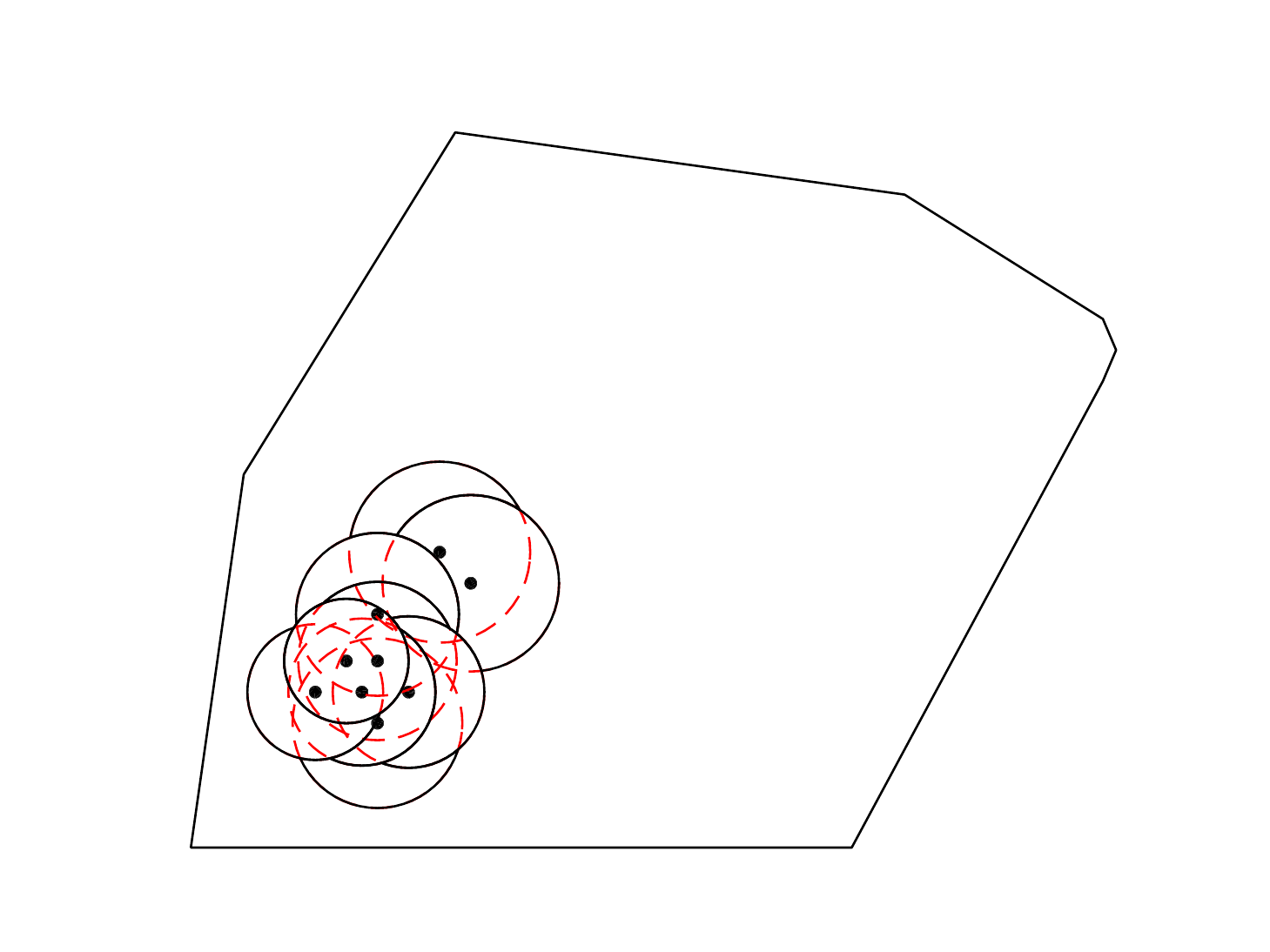}\hspace{0.01cm}
			\includegraphics[width=0.49\textwidth]{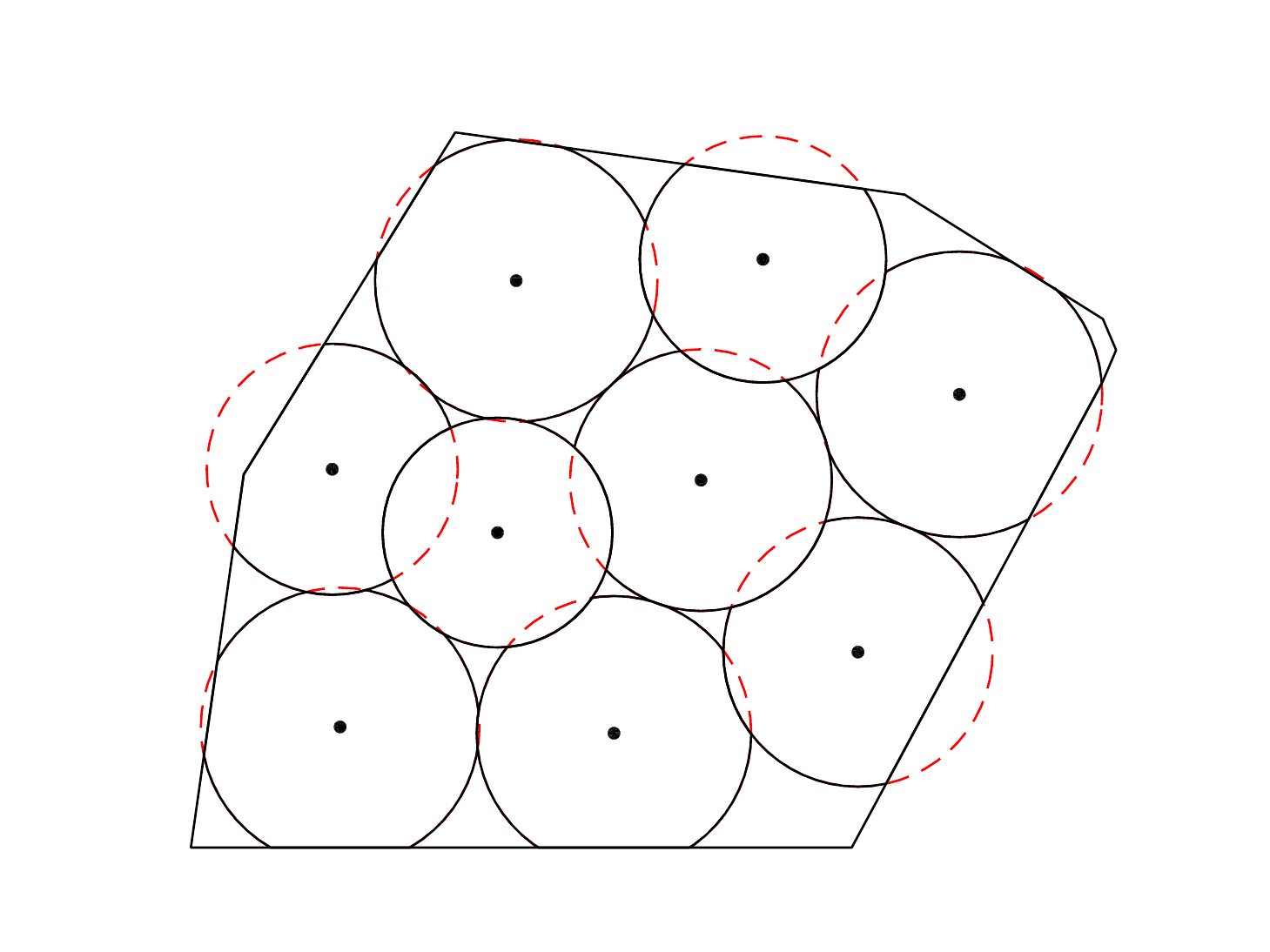}
			\caption{Initial [Left] and final [Right] network configuration and space partitioning.}
			\label{fig:uniform_9_nodes_2D}
		\end{figure}

\begin{figure}[htbp]
			\centering
			\includegraphics[width=0.49\textwidth]{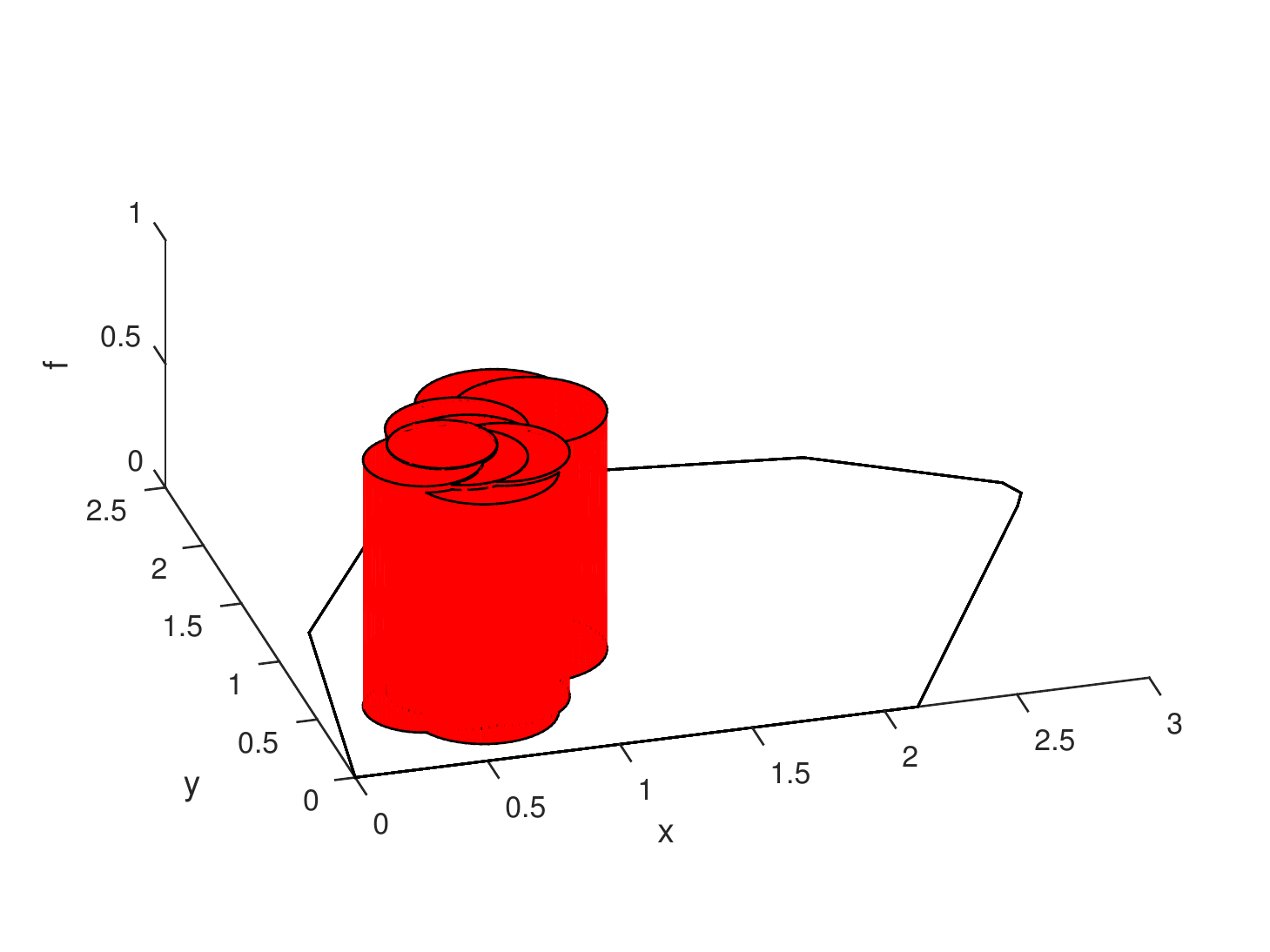}\hspace{0.01cm}
			\includegraphics[width=0.49\textwidth]{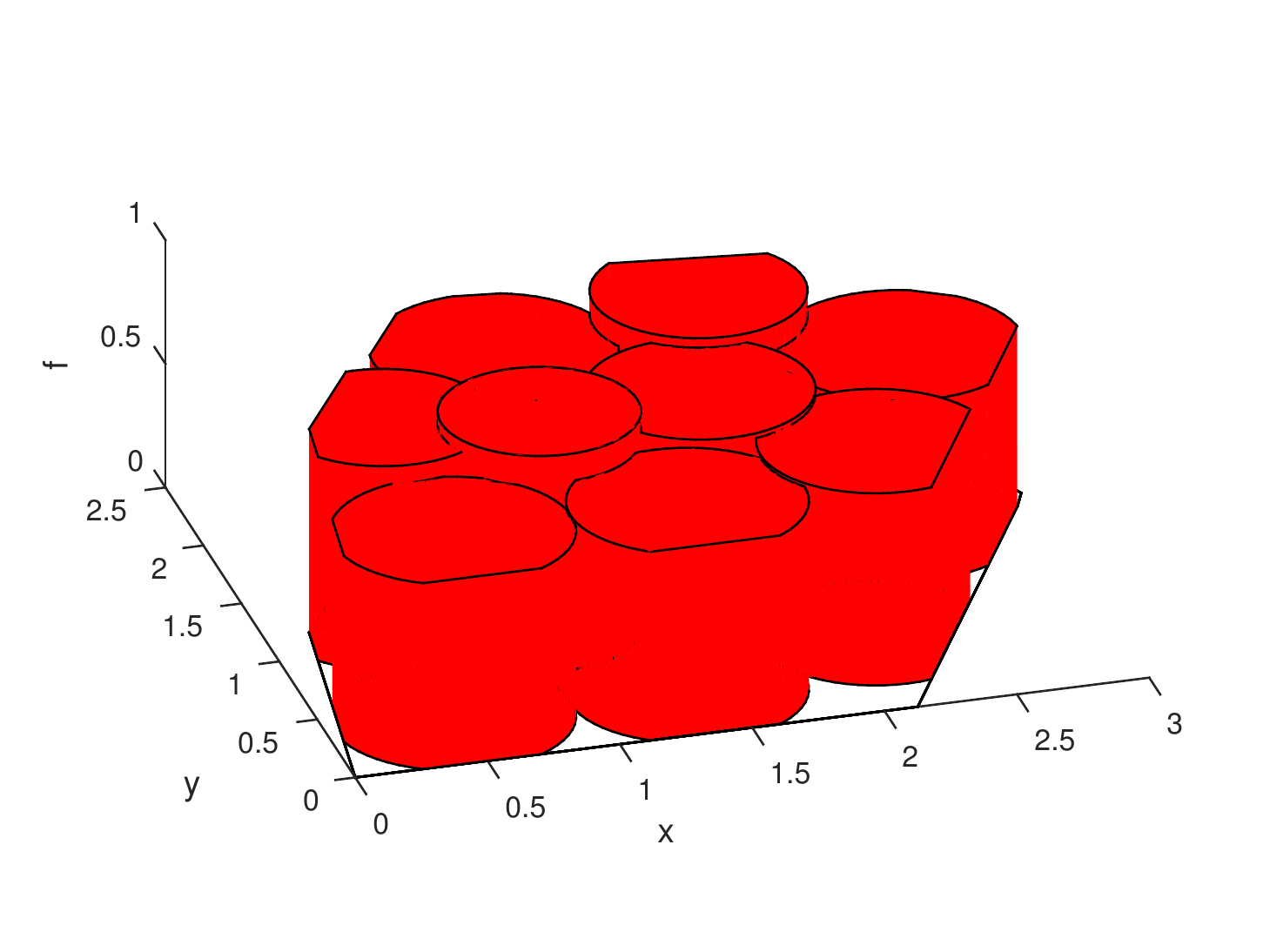}
			\caption{Initial [Left] and final [Right] coverage quality.}
			\label{fig:uniform_9_nodes_3D}
		\end{figure}

\begin{figure}[htbp]
			\centering
			\includegraphics[width=0.8\textwidth]{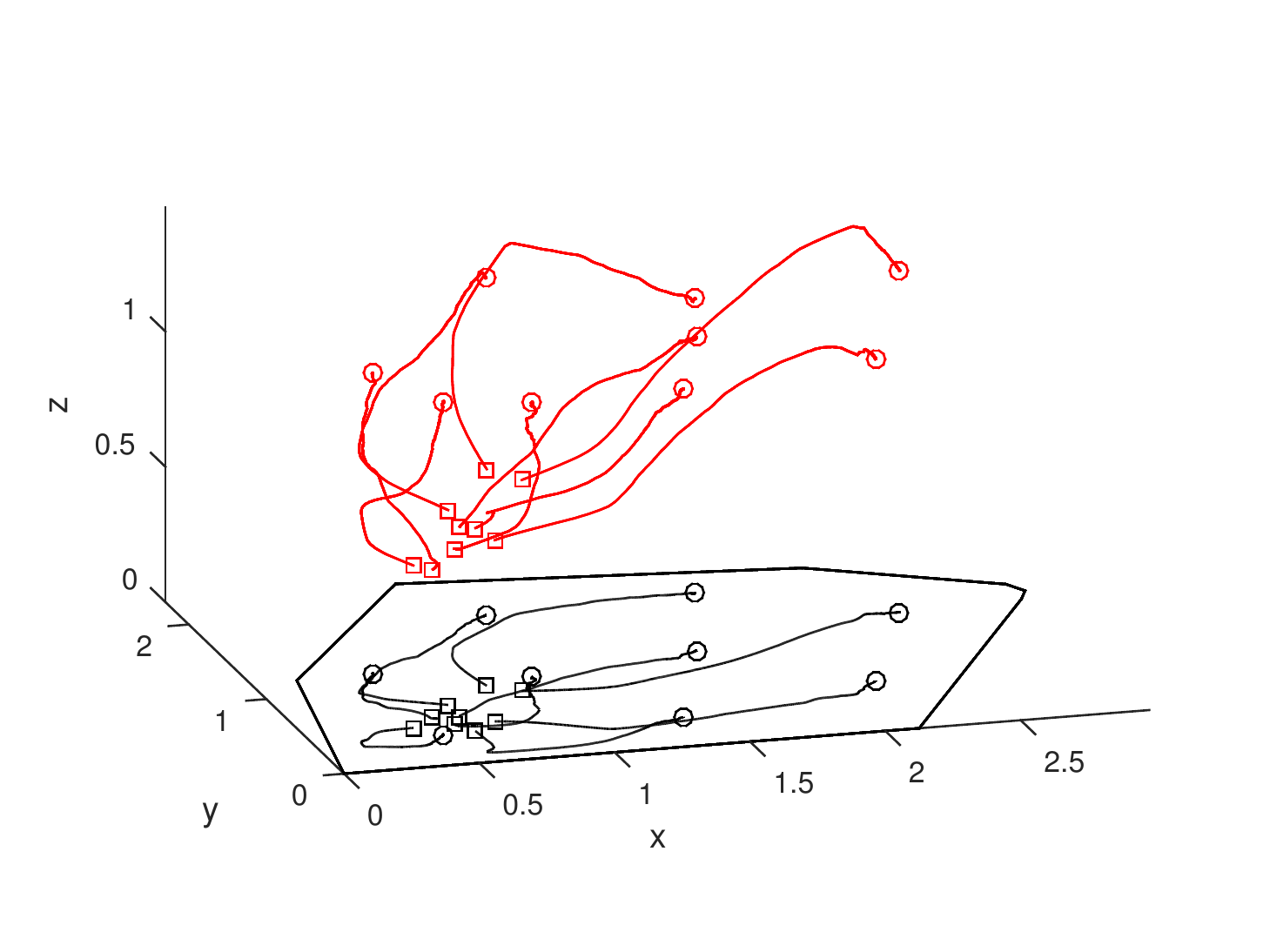}	
			\caption{Node trajectories (blue) and their projections on the sensed region (black).}
			\label{fig:uniform_9_nodes_traj}
		\end{figure}

\begin{figure}[htbp]
			\centering
			\includegraphics[width=0.49\textwidth]{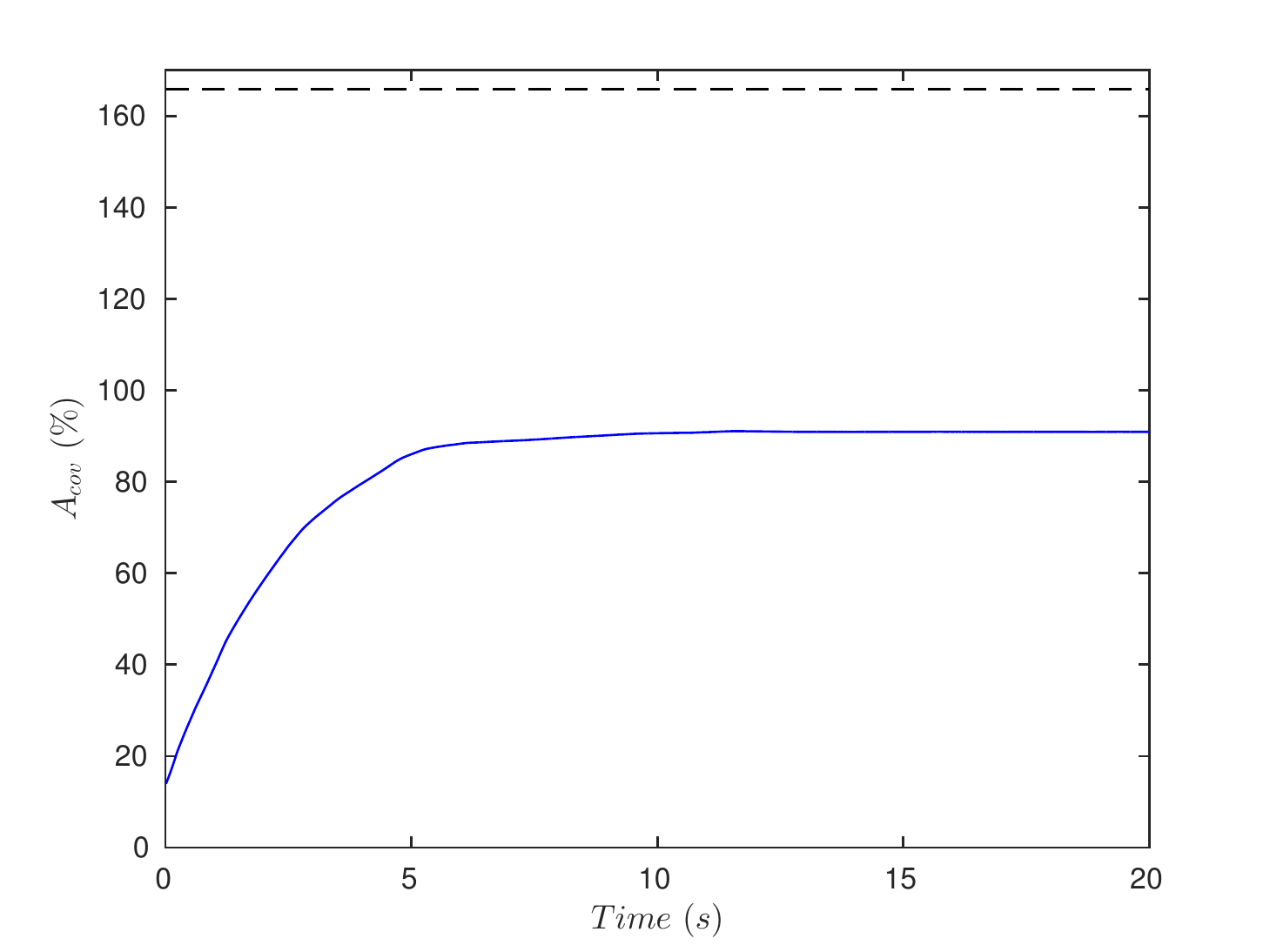}\hspace{0.01cm}
			\includegraphics[width=0.49\textwidth]{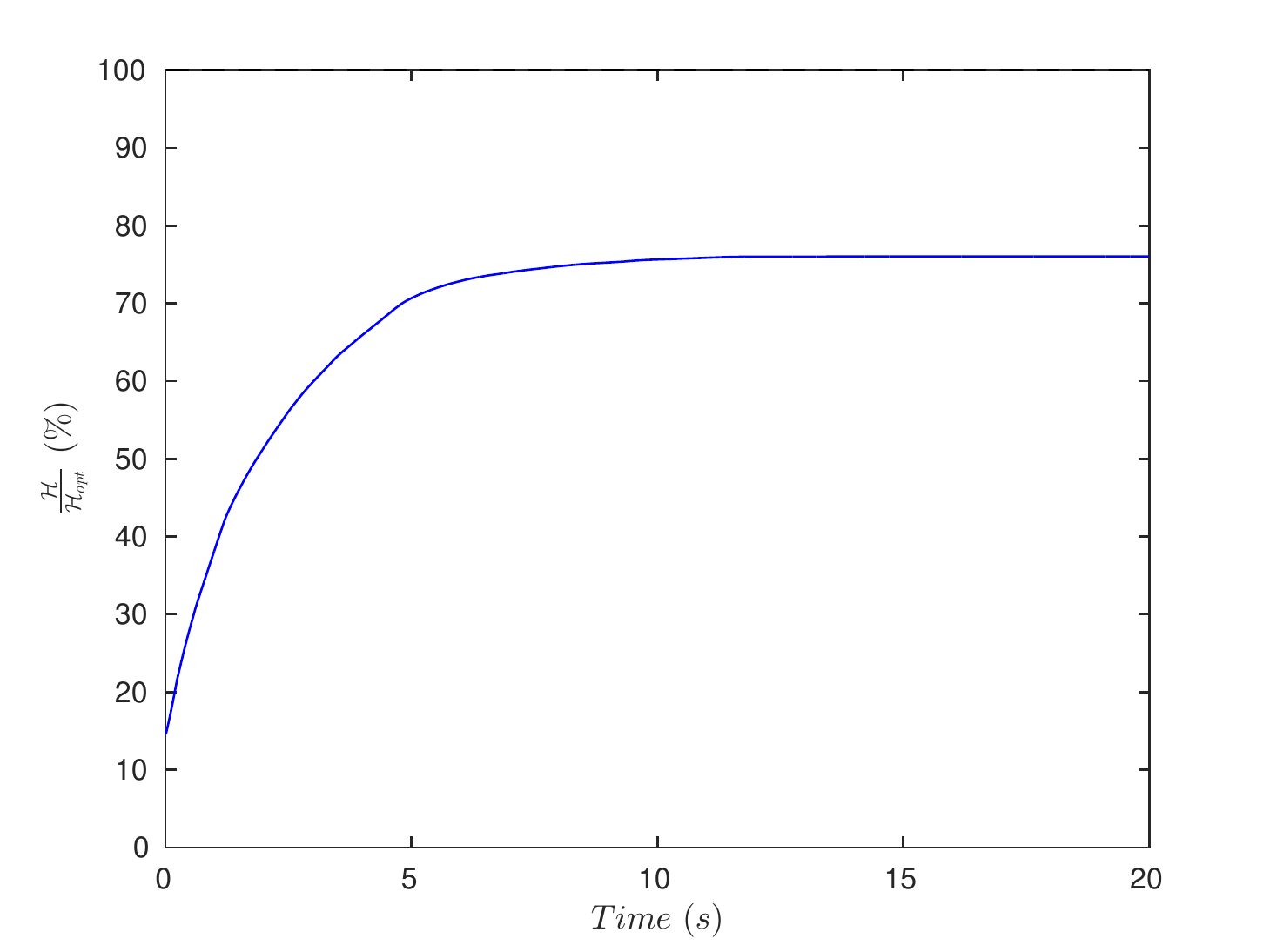}
		\caption{$\frac{\mathcal{A}\left( \bigcup_{i \in I_n} C_i^s \right)}{\mathcal{A}(\Omega)}$ [Left] and $\frac{\mathcal{H}}{\mathcal{H}_{opt}}$ [Right].}
			\label{fig:uniform_9_nodes_area}
		\end{figure}

	A network of nine nodes, identical to those in Case Study I, is examined in this simulation with an initial configuration as seen in Figure \ref{fig:uniform_9_nodes_2D} [Left]. The region $\Omega$ is not large enough to contain these nine $C_{i,opt}^s$ disks and so the nodes converge at different altitudes below $z^{opt}$. This is why the covered area never reaches $\mathcal{A}\left( \bigcup_{i \in I_n} C_{i,opt}^s \right)$, which is larger than $\mathcal{A}(\Omega)$ and why $\mathcal{H}$ never reaches $\mathcal{H}_{opt}$, as seen in Figure \ref{fig:uniform_9_nodes_area}. It can be clearly seen though from Figure \ref{fig:uniform_9_nodes_2D} [Right] and Figure \ref{fig:uniform_9_nodes_area} [Left] that the network covers a significant portion of $\Omega$ with better quality than Case Study I. The volume of the cylinders in Figure \ref{fig:uniform_9_nodes_3D} [Right] has reached a local optimum. The trajectories of the  MAAs in $\mathbb{R}^3$ can be seen in Figure \ref{fig:uniform_9_nodes_traj} in red and their projections on the region of interest in black. The initial positions of the  MAAs are marked by squares and their final positions by circles. It can be seen from the trajectories that the altitude of some nodes was not constantly increasing. This is expected behavior since nodes at lower altitude will increase the stable altitude of nodes at higher altitude they share sensed regions with. Once they no longer share sensed regions, or share a smaller portion, the stable altitude of the upper node will decrease, leading to a decrease in their altitude.

	% % % % % % % % % % % % FIGURES 9 NODES % % % % % % % % % % % %
		
%%%%%%%%%%%%%%%%%%%%%%%%%%%%%%%%%%%%%%%%%

\section{Conclusions}
Area coverage by a network of  MAAs has been studied in this article by use of a combined coverage-quality metric. A partitioning scheme based on coverage quality is employed to assign each MAA an area of responsibility. The proposed control law leads the network to a locally optimal configuration which provides a compromise between covered area and coverage quality. It also guarantees that the altitude of all  MAAs will remain within a predefined range, thus avoiding potential obstacles while also keeping the  MAAs below their maximum operational altitude and in range of their base station. Simulation studies are presented to indicate the efficiency of the proposed control algorithm.

\section*{APPENDIX A - Evaluation of Jacobian matrices}
\label{app:A}

The parametric equation of the boundary of the sensing disk $C_{i}^{s}(X_i,a)$ defined in (\ref{sensing}) is
\begin{equation*}
\gamma_i(k) \colon~ \left[\begin{array}{c} x \\ y \end{array}\right] = \left[\begin{array}{c} x_i + z_i \tan(a) ~\cos(k) \\ y_i + z_i \tan(a) ~\sin(k) \end{array}\right], ~k \in [0,2\pi)
\label{circle_parametric}
\end{equation*}

We will first evaluate $n_i$, $\upsilon_i^i(q)$ and $\nu_i^i(q)$ on $\partial W_i \cap \partial\mathcal{O}$ which is always an arc of the circle $\gamma_i(k)$ because of the partitioning scheme (\ref{partitioning}). The normal vector $n_i$ is given by
\begin{equation*}
n_i = \left[\begin{array}{c} \cos(k) \\ \sin(k) \end{array}\right], ~k \in [0,2\pi).
\end{equation*}
It can be shown that
\begin{equation*}
\upsilon_i^i(q) = \left[\begin{array}{cc} \frac{\partial x}{\partial x_i} & \frac{\partial x}{\partial y_i} \\ \frac{\partial y}{\partial x_i} & \frac{\partial y}{\partial y_i} \end{array}\right] = \left[\begin{array}{cc} 1 & 0 \\ 0 & 1 \end{array}\right] = \mathbb{I}_2
\end{equation*}
and similarly that
\begin{equation*}
\nu_i^i(q) = \left[\begin{array}{c} \frac{\partial x}{\partial z_i} \\ \frac{\partial y}{\partial z_i} \end{array}\right] = \left[\begin{array}{c} \tan(a) \cos(k) \\ \tan(a) \sin(k) \end{array}\right], ~k \in [0,2\pi)
\end{equation*}
resulting in
\begin{equation*}
\nu_i^i(q) \cdot n_i = \tan(a).
\end{equation*}

We will now evaluate $n_i$, $\upsilon_i^i(q)$ and $\nu_i^i(q)$ on $\partial W_j \cap \partial W_i$.

If $f(z_i) = f(z_j)$, the evaluation of $n_i$, $\upsilon_i^i(q)$ and $\nu_i^i(q)$ is irrelevant since the corresponding integral will be $0$ due to the $f(z_i) - f(z_j)$ term.

If $f(z_i) > f(z_j)$, then according to the partitioning scheme (\ref{partitioning}), $\partial W_j \cap \partial W_i$ will be an arc of $\gamma_i(k)$. Thus the evaluation of $n_i$, $\upsilon_i^i(q)$ and $\nu_i^i(q)$ is the same as it was over $\partial W_i \cap \partial\mathcal{O}$. 

If $f(z_i) < f(z_j)$, then according to the partitioning scheme (\ref{partitioning}), $\partial W_j \cap \partial W_i$ will be an arc of $\gamma_j(k)$. Thus both $\upsilon_i^i(q)$ and $\nu_i^i(q)$ will be $0$, since $C_{j}^{s}(X_j,a)$ is not dependent on $X_i$.

To sum up, the evaluation of $\upsilon_i^i(q)$ and $\nu_i^i(q)$ over $\partial W_j \cap \partial W_i$ are the following
\begin{eqnarray}
\nonumber
\upsilon_i^i &=& \left \{
\begin{aligned}
	\mathbb{I}_2, ~ z_i < z_j\\
	\textbf{0}_2, ~ z_i \geq z_j
\end{aligned}
\right.\\
\nonumber
\nu_i^i \cdot n_i &=& \left \{
\begin{aligned}
	\tan(a), ~ z_i < z_j\\
	0, ~ z_i \geq z_j
\end{aligned}
\right.
\end{eqnarray}
where
\begin{equation*}
\textbf{0}_2 = \left[\begin{array}{cc} 0 & 0 \\ 0 & 0 \end{array}\right].
\end{equation*}
It is thus concluded that for the integrals over $\partial W_j \cap \partial W_i$ for the control law of node $i$, only arcs where $f(z_i) > f(z_j)$ need to be considered.

\section*{APPENDIX B - Equilibrium points}
\label{app:B}
The dynamical system can be written as
\begin{eqnarray}
\nonumber
\dot{z}_i ~= u_{i,z}^{opt} ~= \pi ~\tan^2(a)~z_i~\left[ 2~f(z_i) ~+ ~z_i ~f_d(z_i) \right].
\end{eqnarray}
Since $f(z_i)$ and $f_d(z_i)$ are 4th and 3rd degree polynomials respectively, the system has five equilibrium points, one of them being
\begin{equation}
z_1^{eq} = 0.
\end{equation}
The other four are the solutions of the 4th degree polynomial $2~f(z_i) ~+ ~z_i ~f_d(z_i) = 0$ whose analytic expressions are
\begin{eqnarray}
\nonumber
z_2^{eq} &=& z^{\max} \\
\nonumber
z_3^{eq} &=& 2~z^{\min} - z^{\max} \\
\nonumber
z_4^{eq} &=& \frac{2}{3}~z^{\min} - \frac{1}{3} \sqrt{Q} \\
\nonumber
z_5^{eq} &=& \frac{2}{3}~z^{\min} + \frac{1}{3} \sqrt{Q}
\end{eqnarray}
where $Q$ is defined in (\ref{Q_P_definition}), thus all equilibrium points are real. 

We will examine which of these equilibrium points reside in the interval $D = [z^{\min}, ~z^{\max}]$.

Equilibrium point $z_1^{eq} = 0 \notin D$ since $z^{\min} > 0$.

Equilibrium point $z_2^{eq} = z^{\max} \in D$.

Equilibrium point $z_3^{eq} = 2~z^{\min} - z^{\max} < z^{\min}$ thus $z_3^{eq} \notin D$.

Equilibrium point $z_4^{eq} = \frac{2}{3}~z^{\min} - \frac{1}{3} \sqrt{Q} < z^{\min}$ thus $z_4^{eq} \notin D$.

Equilibrium point $z_5^{eq} = \frac{2}{3}~z^{\min} + \frac{1}{3} \sqrt{Q} \in D$ since $z_5^{eq} > z^{\min}$ and $z_5^{eq} < z^{\max}$.

Thus the only equilibrium points in the interval $[z^{\min}, ~z^{\max}]$ are
\begin{eqnarray}
\nonumber
z_2^{eq} &=& z^{\max} \\
\nonumber
z_5^{eq} &=& \frac{2}{3}~z^{\min} + \frac{1}{3} \sqrt{Q}.
\end{eqnarray}

\section*{APPENDIX C - Sign of $u_{i,z}^{opt}$}
\label{app:C}
Since $u_{i,z}^{opt}(z_i)$ is a fifth degree polynomial function, thus both $u_{i,z}^{opt}$ and its derivative $\frac{\partial u_{i,z}^{opt}}{\partial z_i}$ are continuous functions. As a result the sign of $u_{i,z}^{opt}$ will be constant between consecutive roots of $u_{i,z}^{opt} = 0$. Since we are interested in the sign of $u_{i,z}^{opt}$ in the interval $[z^{\min}, ~z^{\max}]$ and the only roots in that interval are $z^{\max}$ and $z_5^{eq} \in (z^{\min}, ~z^{\max})$, as shown in Appendix B, we just need to evaluate the sign of $u_{i,z}^{opt}$ in the intervals $\left[z^{\min}, z_5^{eq}\right)$ and $\left(z_5^{eq}, z^{\max}\right)$.

We will show that $u_{i,z}^{opt} > 0, ~~~\forall z_i \in \left[z^{\min}, z_5^{eq}\right)$ by substituting $z_i^p = \frac{z^{\min}+z_5^{eq}}{2}$ into $u_{i,z}^{opt}$. After tedious algebraic manipulations it can be shown that the inequality $u_{i,z}^{opt}(z_i^p) > 0$ is equivalent to
\footnotesize
\begin{eqnarray}
\nonumber
\left(9 {z^{\max}}^2 - 18 z^{\max} z^{\min} +11 {z^{\min}}^2 - 2 z^{\min} \sqrt{Q} \right) 
\left(33 {z^{\max}}^2 -66 z^{\max} z^{\min} +31 {z^{\min}}^2 + 2 z^{\min} \sqrt{Q} \right) &>& 0 \Rightarrow \\
\nonumber
\left(3P - R \right) \cdot \left(11P + R\right) &>& 0
\end{eqnarray}
\normalsize
where $R \stackrel{\triangle}{=} 2 z^{\min} \sqrt{Q} - 2 {z^{\min}}^2$.
Since $R > 0$ and $P > 0$ from (\ref{Q_P_definition}), we have that $11P+R > 0$ and after tedious algebraic manipulations it can be shown that $3P - R > 0$ since
\begin{eqnarray}
\nonumber
3P - R = 9 {z^{\max}}^2 - 18 z^{\max} z^{\min} + 9 {z^{\min}}^2 - 2 z^{\min} \sqrt{Q} + 2 {z^{\min}}^2 &>& 0 \Rightarrow \\
\nonumber
\left( 9 {z^{\max}}^2 - 18 z^{\max} z^{\min} + 11 {z^{\min}}^2 \right)^2 &>& 4 {z^{\min}}^2 Q.
\end{eqnarray}
Substitution of $Q$ from (\ref{Q_P_definition}) yields
\begin{eqnarray}
\nonumber
27 P + 8 {z^{\min}}^2 &>& 0
\end{eqnarray}
Thus it is proven that $u_{i,z}^{opt}(z_i^p) > 0$ and consequently that $u_{i,z}^{opt} > 0, ~~~\forall z_i \in [z^{\min}, z_5^{eq}]$.

We will show that $u_{i,z}^{opt} < 0, ~~~\forall z_i \in \left(z_5^{eq}, z^{\max}\right)$ by evaluating the derivative of $u_{i,z}^{opt}$ at $z^{\max}$
\begin{equation*}
\frac{\partial u_{i,z}^{opt}}{\partial z_i}(z^{\max}) = \frac{8 \pi (\tan a)^2 {z^{\max}}^2}{\left(z^{\max} - z^{\min}\right)^2} > 0.
\end{equation*}
Hence $u_{i,z}^{opt}(z^{\max}) = 0$ and $\frac{\partial u_{i,z}^{opt}}{\partial z_i}(z^{\max}) > 0$. 

Since $\frac{\partial u_{i,z}^{opt}}{\partial z_i}$ is a continuous function and $\frac{\partial u_{i,z}^{opt}}{\partial z_i}(z^{\max}) > 0$, there is a region $E$ around $z^{\max}$ inside which $\frac{\partial u_{i,z}^{opt}}{\partial z_i} > 0$. Thus $z^{\max}-\epsilon \in E$ and $\frac{\partial u_{i,z}^{opt}}{\partial z_i} > 0, ~\forall z_i \in \left[z^{\max}-\epsilon,~z^{\max}\right]$, where $\epsilon$ is an infinitesimally small positive constant. Since $u_{i,z}^{opt}$ is an increasing function in the interval $\left[z^{\max}-\epsilon,~z^{\max}\right]$, it is true that
\begin{eqnarray}
u_{i,z}^{opt}(z^{\max}-\epsilon) &<& 0
\end{eqnarray}
%We will now evaluate the derivative at $z^{\max}-\epsilon$, where $\epsilon$ is an infinitesimally small positive constant. Since $\frac{\partial u_{i,z}^{opt}}{\partial z_i}(z^{\max}) > 0$ and $\frac{\partial u_{i,z}^{opt}}{\partial z_i}$ is a continuous function, it is also true that $\frac{\partial u_{i,z}^{opt}}{\partial z_i}(z^{\max}-\epsilon) > 0$.
%
%Since the derivative $z^{\max}-\epsilon$ is positive, it is true that
Since the sign of $u_{i,z}^{opt}$ is constant in the interval $\left(z_5^{eq}, z^{\max}\right)$, we obtain that $u_{i,z}^{opt} < 0, ~~~\forall z_i \in \left(z_5^{eq}, z^{\max}\right)$.

\section*{References}

\bibliography{bibliography/ysphdbook,bibliography/SP_bibliography}

\end{document}